\documentclass[12pt]{article}
\pdfoutput=1
\usepackage{amsmath,amssymb,amsthm,amsxtra,overpic,bbm,bm,epsfig,subfigure}
\usepackage{mathrsfs}
\usepackage{graphicx}
\usepackage{epstopdf}
\usepackage{color}
\usepackage{comment}
\usepackage{float}
\usepackage{cite}
\usepackage[utf8]{inputenc}
\usepackage{slashed,stmaryrd}

\begin{document}

\vspace{0.2cm}
\begin{center}
{\Large\bf Constraining Neutrino Lifetimes and Magnetic Moments via Solar Neutrinos in the Large Xenon Detectors}
\end{center}
\vspace{0.2cm}

\begin{center}
{\bf Guo-yuan Huang~$^{a,~b}$}~\footnote{E-mail: huanggy@ihep.ac.cn},
\quad
{\bf Shun Zhou~$^{a,~b}$}~\footnote{E-mail: zhoush@ihep.ac.cn}
\\
\vspace{0.2cm}
{\small $^a$Institute of High Energy Physics, Chinese Academy of
Sciences, Beijing 100049, China \\
$^b$School of Physical Sciences, University of Chinese Academy of Sciences, Beijing 100049, China}
\end{center}

\vspace{1.5cm}

\begin{abstract}
The multi-ton-scale liquid xenon detectors, with an excellent energy resolution of a few keV, will be constructed to probe the dark-matter particles. In this paper, we show that precision measurements of the low-energy solar neutrinos via the elastic neutrino-electron scattering in this kind of detectors are able to improve the present limits on neutrino lifetimes and neutrino magnetic moments by about one order of magnitude. We carefully study the impact of the unknown neutrino mass spectrum on the ultimate limits in the case of non-radiative visible neutrino decays. In the case of invisible neutrino decays, the lower bounds $\tau^{}_1/m^{}_1 \gtrsim 3 \times 10^{-2}~{\rm s}/{\rm eV}$ and $\tau^{}_2/m^{}_2 \gtrsim 8\times 10^{-3}~{\rm s}/{\rm eV}$ at the $2\sigma$ level can be obtained for a total exposure of $70~{\rm ton}\cdot {\rm year}$. Furthermore, a restrictive constraint on the effective magnetic moment of neutrinos $\mu^{}_{\rm eff} \lesssim 2.6\times 10^{-12}~\mu^{}_{\rm B}$, with $\mu^{}_{\rm B}$ being the Bohr magneton, can be achieved. This is among the best results that will be available in the laboratory experiments in the near future.
\end{abstract}

\newpage

\section{Introduction}

That neutrinos are massive particles has been firmly established by a great number of neutrino oscillation experiments, but it remains to know whether the neutrino mass ordering is normal (i.e., $m^{}_1 < m^{}_2 < m^{}_3$) or inverted (i.e., $m^{}_3 < m^{}_1 < m^{}_2$) and whether massive neutrinos are Dirac or Majorana particles~\cite{Akhmedov:1999uz,GonzalezGarcia:2002dz,Strumia:2006db,GonzalezGarcia:2007ib,Xing:2011zza}. Besides the neutrino mass ordering and the Majorana nature, other intrinsic properties of massive neutrinos are also very important for exploring the origin of neutrino masses and the underlying new physics beyond the Standard Model (SM). For instance, it is quite intriguing to probe how long massive neutrinos can live and how large their electromagnetic dipole moments can be~\cite{Giunti:2014ixa,Balantekin:2018azf}. In the minimal extension of the SM with nonzero neutrino masses, the rate of the radiative neutrino decay $\nu^{}_i \to \nu^{}_j + \gamma$ is given by~\cite{Fujikawa:1980yx, Shrock:1982sc, Xing:2012gd}
\begin{eqnarray}\label{eq:decayrate}
\Gamma^{}_{ij} = \frac{(m^2_i - m^2_j)^3}{8\pi m^3_i} \left(|\mu^{}_{ij}|^2 + |\epsilon^{}_{ij}|^2\right) \approx 5.3\times \left(1 - \frac{m^2_j}{m^2_i}\right)^3 \left(\frac{m^{}_i}{1~{\rm eV}}\right)^3 \left(\frac{\mu^{}_{\rm eff}}{\mu^{}_{\rm B}}\right)^2~{\rm s}^{-1} \; ,
\end{eqnarray}
where $\mu^{}_{\rm eff} \equiv \left(|\mu^{}_{ij}|^2 + |\epsilon^{}_{ij}|^2\right)^{1/2}$ stands for the effective neutrino magnetic moment, $\mu^{}_{ij}$ and $\epsilon^{}_{ij}$ for the transitional magnetic and electric dipole moments, and $\mu^{}_{\rm B} \equiv e/(2m^{}_e)$ for the Bohr magneton. Note that the neutrino masses $m^{}_i$ and $m^{}_j$ involved in the radiative decay $\nu^{}_i \to \nu^{}_j + \gamma$ should satisfy $m^{}_i > m^{}_j$, otherwise such a decay process is kinematically forbidden. Given the latest experimental data on neutrino mass-squared differences and flavor mixing angles~\cite{Tanabashi:2018oca}, one can obtain $\mu^{}_{\rm eff} \lesssim 10^{-24}~\mu^{}_{\rm B}$ in the limit that the lightest neutrino is massless~\cite{Xing:2012gd}. Therefore, the corresponding neutrino lifetime $\tau^{}_i \equiv \Gamma^{-1}_i$, where the total rate is $\Gamma^{}_i \equiv \sum_j \Gamma^{}_{ij}$ (for $j\neq i$), can be estimated as $\tau^{}_i \gtrsim 10^{43}$ years, which is much longer than the age of our Universe.

Since the effective neutrino magnetic moment $\mu^{}_{\rm eff}$ and the neutrino decay rate $\Gamma^{}_i$ are so small in the SM with nonzero neutrino masses, any new-physics contributions that enhance $\mu^{}_{\rm eff}$ and $\Gamma^{}_i$ to the level within the reach of current and future experiments will receive severe constraints. It has been shown in Ref.~\cite{Xing:2012gd} that $\mu^{}_{\rm eff}$ and $\Gamma^{}_i$ can be increased at most by a factor of $10^2$ and $10^4$, respectively, in the canonical seesaw model, due to the non-unitarity of the flavor mixing matrix of light neutrinos. In other new-physics models~\cite{Chikashige:1980qk, Schechter:1981cv, Gelmini:1982rr, Zhou:2007zq, Berezhiani:1989fp,Sakharov:1994pr}, the non-radiative neutrino decays $\nu^{}_i \to \nu^{}_j + \eta$ with $\eta$ being an invisible particle can take place and the corresponding decay rates can be remarkably large. As the underlying new physics associated with neutrino masses is not yet known, a phenomenological approach is usually adopted and proves to be useful. Namely, taking the lifetime $\tau^{}_i$ for the neutrino mass eigenstate $\nu^{}_i$ of mass $m^{}_i$ to be a free parameter, one can make use of all the available experimental data to draw a robust limit on $\tau^{}_i$ for a given neutrino mass or on the combination $\tau^{}_i/m^{}_i$. It is evident that such limits can then be applied to any particle-physics models that are constructed to explain neutrino masses and predicting significant decay rates of massive neutrinos.

In fact, there exist a variety of constraints on the neutrino lifetimes from the observational data, which can be summarized in three distinct categories:
\begin{itemize}
\item Radiative Decays (RDs) -- The rate for the RD of a massive neutrino $\nu^{}_i \to \nu^{}_j + \gamma$ has been given in Eq.~(\ref{eq:decayrate}), and the final-state neutrino or photon will be visible in the neutrino detectors or optical telescopes. In practice, the observational constraints on the RDs of massive neutrinos can be translated into those on the effective neutrino magnetic moment $\mu^{}_{\rm eff}$ and vice versa, as clearly indicated in Eq.~(\ref{eq:decayrate}). The RDs of massive neutrinos in the Universe lead to the spectral distortion of the cosmic microwave background (CMB), which is well known to be a nearly perfect black-body radiation. See, e.g., Ref.~\cite{Mirizzi:2007jd} for an earlier study in this connection. A recent analysis of the CMB spectral data measured with COBE-FIRAS~\cite{Aalberts:2018obr} yields the lower limits on the neutrino lifetime $\tau^{}_{21} \gtrsim 10^{21}~{\rm s}$ if the small neutrino mass-squared difference $\Delta m^2_{21} \approx 7.5\times 10^{-5}~{\rm eV}^2$ is considered, or $\tau^{}_{32} \sim \tau^{}_{31} \gtrsim 10^{19}~{\rm s}$ for the large neutrino mass-squared difference $|\Delta m^2_{31}| \approx 2.5\times 10^{-3}~{\rm eV}^2$, at the $95\%$ confidence level (C.L.). However, these limits correspond to the upper bound on the effective neutrino magnetic moment $\mu^{}_{\rm eff} \lesssim (10^{-8} \cdots 10^{-7})~\mu^{}_{\rm B}$, which is weaker by about four orders of magnitude than the astrophysical bounds $\mu^{}_{\rm eff} \lesssim 3\times 10^{-12}~\mu^{}_{\rm B}$~\cite{Raffelt:1999gv}. See Refs.~\cite{Giunti:2014ixa, Giunti:2015gga} for recent reviews on the astrophysical and laboratory bounds on neutrino magnetic moments.

\item Non-Radiative Visible Decays (NRVDs) --  This type of decays $\nu^{}_i \to \nu^{}_j + \eta$, where the new particle $\eta$ has been safely assumed to be immune from any SM interactions, can take place in a number of neutrino mass models, such as the singlet majoron model\cite{Chikashige:1980qk, Schechter:1981cv, Gelmini:1982rr}, and in other new-physics scenarios~\cite{Zhou:2007zq}. However, the final-state neutrino can in principle be observed in neutrino detectors.

\item Invisible Decays (IVDs) -- Massive neutrinos can also experience the IVDs into the particles all in the hidden sector, implying that neutrino mass eigenstates simply disappear after their production. Obviously, the complete or partial disappearance of neutrinos in various experimental and astrophysical environments will cause contradictions with the existing neutrino observations, leading to restrictive constraints on neutrino lifetimes.
\end{itemize}
For both cases of NRVDs and IVDs, many different bounds on neutrino lifetimes have been drawn from cosmological and astrophysical observations. First, the cosmological data on the CMB anisotropies are favoring free-streaming neutrinos, which can be implemented to constrain the interaction between neutrinos $\nu^{}_i$ and the scalar particle $\eta$, leading to a restrictive bound $\tau^{}_i/m^{}_i \gtrsim 4 \times 10^{11}~m^2_{50}~{\rm s}/{\rm eV} $ on neutrino lifetimes~\cite{Hannestad:2005ex} with $m^{}_{50} \equiv m^{}_i/(50~{\rm meV})$. It has been further pointed out in Ref.~\cite{Serpico:2007pt} that the precise determination of absolute neutrino masses from the cosmological observations could be the best probe of neutrino lifetimes, i.e., $\tau^{}_i/m^{}_i \gtrsim 10^{16}~m^{-5/2}_{50}~{\rm s}/{\rm eV}$. This has been derived by requiring that massive neutrinos should survive the time when they become non-relativistic at the redshift of $z^{}_{\rm nr} \approx 100~{m^{}_{50}}$. Second, the neutrino observation of Supernova 1987A indicates at least one mass eigenstate of neutrinos should be stable over the distance to the Earth, placing the lower bound $\tau^{}_i/m^{}_i \gtrsim 5.5 \times 10^{5}~{\rm s}/{\rm eV}$~\cite{Frieman:1987as}.

In this work, we attempt to constrain neutrino lifetimes and magnetic moments via the observations of low-energy solar neutrinos in the future large xenon detectors. Such a study is interesting and well motivated. First, thanks to recent tremendous developments in the direct detection of dark-matter particles, the multi-ton-scale detectors of liquid xenon will be constructed~\cite{Liu:2017drf}. These large detectors are sensitive to the recoil energies of nuclei or electrons in the range of a few keV, so they could also serve as powerful observatories to measure low-energy solar neutrinos via the elastic neutrino-electron scattering. Second, solar neutrino observations have already been utilized to place meaningful bounds on neutrino lifetimes~\cite{Beacom:2002cb, Joshipura:2002fb, Bandyopadhyay:2002qg, Berryman:2014qha, Picoreti:2015ika}. The most restrictive constraints on the IVDs of $\nu^{}_1$ and $\nu^{}_2$ have been obtained with the combined analysis of the latest Borexino data \cite{Bellini:2014uqa, Bellini:2011rx}, KamLAND \cite{Gando:2010aa} and DayaBay \cite{An:2013zwz}, namely, $\tau^{}_{1}/m^{}_{1} \gtrsim 4 \times 10^{-3}~{\rm s}/{\rm eV}$~\cite{Berryman:2014qha}
\footnote{It is worth mentioning that the limits in Ref.~\cite{Berryman:2014qha} are given for the parameters $d^{}_i \equiv m^{}_i/\tau^{}_i$ in units of ${\rm eV}^2$, which have now been converted into those for $\tau^{}_i/m^{}_i$ in units of ${\rm s}/{\rm eV}$.}
 and $\tau^{}_{2}/m^{}_{2} \gtrsim  10^{-3}~{\rm s}/{\rm eV}$~\cite{Picoreti:2015ika} at the $95\%$ C.L.
 We show that the detection of keV solar neutrinos will improve the present constraints on neutrino decays by one order of magnitude, mainly due to the advantages of a low energy threshold and low backgrounds achieved in the future xenon detectors. Other constraints from reactor neutrino experiments~\cite{Abrahao:2015rba}, atmospheric neutrino and long-baseline accelerator neutrino experiments~\cite{GonzalezGarcia:2008ru, Gomes:2014yua, Coloma:2017zpg, Choubey:2018cfz, Choubey:2017eyg, Choubey:2017dyu,Gago:2017zzy,Ascencio-Sosa:2018lbk,deSalas:2018kri} have been extensively discussed in the literature, while the impact of neutrino decays on ultrahigh-energy cosmic neutrinos in neutrino telescopes has been investigated in Refs.~\cite{Beacom:2002vi, Baerwald:2012kc, Dorame:2013lka, Bustamante:2015waa, Bustamante:2016ciw, Denton:2018aml}. Third, if the elastic neutrino-electron scattering receives an extra contribution from the effective neutrino magnetic moment, we are able to get the upper bound $\mu^{}_{\rm eff} \lesssim 2.6 \times 10^{-12}~\mu^{}_{\rm B}$ at the $95\%$ C.L., which is among the best results that can be reached in future laboratory experiments~\cite{Giunti:2015gga}.

The remaining part of this paper is organized as follows. In Section 2, we briefly explain how to observe solar neutrinos in the liquid xenon detectors, and take the DARWIN experiment as a typical example. Section 3 and Section 4 will be devoted to the studies of neutrino decays and neutrino magnetic moments, respectively, where the expected sensitivities of a DARWIN-like experiment will be given. Finally, we summarize our main results in Section 5.

\section{Solar Neutrinos in the Xenon Detectors}

The identity and nature of the dark matter particles have been a long-standing puzzle in particle physics, astrophysics and cosmology~\cite{Bertone:2004pz}. Recent years have seen tremendous progress in the direct experimental searches for the weakly interacting massive particles (WIMPs), which are currently one of the best motivated candidates for the dark matter. As was first suggested in Ref.~\cite{Goodman:1984dc}, the WIMPs can be observed via the recoil energies of heavy nuclei when the WIMP-nucleon scattering occurs in the detector. The existing xenon detectors have already set stringent bounds on the WIMP-nucleon scattering cross section, including LUX~\cite{Akerib:2016vxi, Akerib:2017kat}, PandaX-II~\cite{Tan:2016zwf, Cui:2017nnn}, XENON-100~\cite{Aprile:2016swn} and its ton-scale successor XENON-1T~\cite{Aprile:2017iyp}. The future experiments will hopefully be able to reach the so-called \emph{neutrino floor}, which is composed of the irreducible backgrounds arising from solar neutrinos, atmospheric neutrinos and diffuse supernova neutrinos. Therefore, the ongoing ton-scale and forthcoming multi-ton-scale xenon experiments, like XENON1T \cite{Aprile:2017iyp}, XENONnT \cite{Aprile:2015uzo}, PandaX-4T~\cite{Zhang:2018xdp}, LZ~\cite{Akerib:2015cja}, and DARWIN~\cite{Schumann:2015cpa, Baudis:2013qla, Aalbers:2016jon}, will be very efficient in detecting low-energy neutrinos.

In general, there exist two kinds of observables in the xenon detectors, namely, the nuclear recoil (NR) events and the electron recoil (ER) events. The NR events are induced by the WIMP-nucleon scattering, while the ER events from the WIMP-electron scattering would be one of the backgrounds for the former when they cannot be fully discriminated. The neutrinos can result in both the NR and the ER backgrounds \cite{Billard:2013qya}. According to the energy threshold and the target mass of the xenon detector, it can be verified that the NR neutrino background is basically caused by those neutrinos with energies around or higher than $1~{\rm MeV}$, and the ER neutrino background is mainly contributed by the low-energy solar neutrinos. As the impact of neutrino decays and magnetic moments will be more significant for lower-energy neutrinos, we concentrate on the ER events induced by solar neutrinos on the xenon detectors and study the experimental sensitivities to neutrino lifetimes and magnetic moments.

In the following discussions, we shall take the DARWIN experiment~\cite{Aalbers:2016jon}, which is capable of measuring the flux of solar $pp$ neutrinos to the $1\%$ level, as a typical example. In Ref.~\cite{Baudis:2013qla}, the physics potential of the DARWIN experiment for solar neutrinos has been discussed. However, the expected sensitivities to neutrino lifetimes and magnetic moments have not yet been fully explored. For the detection of low-energy solar neutrinos, the energy window of $(2 \cdots 30)~{\rm keV}$ is accessible for the ER events in the DARWIN detector. In this energy range, the major components of solar neutrinos are the $pp$ neutrinos ($91\%$) and the $^7{\rm Be}$ neutrinos ($7.1\%$). For the latter component with a total flux $\phi^{(^7{\rm Be})}_{\rm tot}$, neutrinos are approximately mono-energetic at $384~{\rm keV}$ ($10.4\%$)  and $862~{\rm keV}$ ($89.6\%$) with a negligible thermal broadening of ${\rm keV}$. Therefore, the spectrum of the $^7{\rm Be}$ neutrinos can be well described by
\begin{align}\label{eq:Be7flux}
\frac{{\rm d} \phi^{(^7{\rm Be})}_{}(E^{}_{\nu})}{{\rm d} E^{}_{\nu}} = \phi^{(^7{\rm Be})}_{\rm tot} \left[\delta(E^{}_{\nu}-384~{\rm keV})\cdot  10.4\% +\delta(E^{}_{\nu}-862~{\rm keV}) \cdot  89.6\% \right] \; .
\end{align}
The spectrum of $pp$ neutrinos is well approximated as~\cite{Raffelt:1996wa, Wurm:2017cmm}
\begin{align}\label{eq:ppflux}
\frac{{\rm d} \phi^{(\rm pp)}_{}(E^{}_{\nu})}{{\rm d} E^{}_{\nu}} = A \left(Q + m^{}_e - E^{}_{\nu}\right) \left[\left(Q + m^{}_e - E^{}_{\nu}\right)^2 - m_e^2\right]^{1/2} E_{\nu}^2 F \; ,
\end{align}
where $Q = 420~{\rm keV}$ is the maximum energy for the neutrino, $F$ is the Fermi function describing the Coulomb interaction of the final-state positron, $A$ is the normalization constant for the neutrino flux. In this work, the solar neutrino fluxes and the corresponding errors are taken from Table 6 of Ref.~\cite{Vinyoles:2016djt}, where the latest nuclear reaction rates have been adopted and a new generation of solar models have been constructed. For illustration, we choose the standard solar model B16-AGSS09met, in which the total fluxes $\phi^{(\rm pp)}_{\rm tot}=6.03~(1\pm 0.005)\times 10^{10}~{\rm cm^{-2}s^{-1}}$ and $\phi^{(\rm {}^7\!Be)}_{\rm tot}=4.5~(1\pm 0.06)\times 10^{9}~{\rm cm^{-2}s^{-1}}$ are predicted. In fact, the real spectrum will be further modified by thermal broadening and screening effects. However, we have checked that such corrections to the spectral shape are less than $1\%$ and thus irrelevant for our discussions.

When solar neutrinos enter into the detectors, the most important and relevant interaction will be the elastic scattering between neutrinos and  electrons therein. In the SM, the differential cross section for the elastic neutrino-electron scattering is given by~\cite{tHooft:1971ucy}
\begin{align} \label{eq:crosssection}
\frac{{\rm d} \sigma(T^{}_e, E^{}_{\nu})}{{\rm d} T^{}_e} = \frac{2 G_{\rm F}^2 m^{}_e}{\pi} \left[g_{\rm L}^2 + g_{\rm R}^2 \left(1 - \frac{T^{}_e}{E^{}_{\nu}}\right)^2 - g^{}_{\rm L} g^{}_{\rm R} \frac{m^{}_e T^{}_e}{E_{\nu}^2}\right] \; ,
\end{align}
with $g^{}_{\rm L} = \pm 1/2 + \sin^2 \theta^{}_{\rm W}$ and $g^{}_{\rm R} = \sin^2 \theta^{}_{\rm W}$, where $g^{}_{\rm L}$ takes the plus sign for $\nu^{}_e$ but the negative sign for $\nu^{}_\mu$ and $\nu^{}_\tau$,  $E^{}_{\nu}$ denotes the energy of the initial neutrino, $T^{}_e \equiv E^{}_e - m^{}_e$ is the  recoil energy of the final-state electron. The low-energy value of the weak mixing angle~\cite{Erler:2004in}, i.e., $\sin^2 \theta^{}_{\rm W}=0.23867$
\footnote{The average value of the weak mixing angle from the $Z$-pole measurements is well consistent with the SM prediction~\cite{Kumar:2013yoa}, while some other low energy measurements are not. If we implement the central value and the uncertainty directly from the low energy measurement of the parity nonconservation in Cesium~\cite{Dzuba:2012kx}, i.e. $\sin^2 \theta^{}_{\rm W}=0.2356(20)$, the $\mathcal{O}(1\%)$ error will lead to an $\mathcal{O}(0.5\%)$ uncertainty in the final ER events, which is slightly smaller than the uncertainty caused by the solar neutrino fluxes. The impact of including this error has been found to be negligible for our results of the sensitivities.}
is adopted in our calculations, and its tiny error of $\mathcal{O}(10^{-4})$ can be ignored.

Finally, the event spectrum can be found out by integrating the product of neutrino spectrum in Eq.~(\ref{eq:Be7flux}) or Eq.~(\ref{eq:ppflux}) and the differential cross section in Eq.~(\ref{eq:crosssection}) over the initial neutrino energy, namely,
\begin{align} \label{eq:events}
\frac{{\rm d} N(T^{}_e)}{{\rm d} T^{}_e} = N^{}_0 \cdot t \cdot \sum_{ \alpha = e, \mu, \tau} \int_{E^{\rm min}_{\nu}}^{E^{\rm max}_{\nu}} {\rm d}E^{}_{\nu} \frac{{\rm d} \phi^{}_{\alpha}(E^{}_{\nu})}{{\rm d} E^{}_{\nu}} \cdot \frac{{\rm d} \sigma^{}_{\alpha}(T^{}_e, E^{}_{\nu})}{{\rm d} T^{}_e} \; ,
\end{align}
where $N^{}_0$ is the total number of target electrons in the detector, $t$ is the effective running time, $E^{\rm min}_{\nu} = (1+\sqrt{1+2m^{}_{e}/T^{}_{e}})T^{}_{e}/2$ is the minimum neutrino energy required to have an electron recoil energy of $T^{}_{e}$, and the integration limit $E^{\rm max}_{\nu} = Q$ for the $pp$ neutrinos is the maximum neutrino energy. Here the differential flux ${{\rm d} \phi^{}_{\alpha}(E^{}_{\nu})}/{{\rm d} E^{}_{\nu}}$ for each neutrino flavor at the detector receives contributions from both $pp$ and $^7{\rm Be}$ neutrinos after flavor conversions. In Eq.~(\ref{eq:events}), the ER events induced by all three-flavor neutrinos have been taken into account. As is well known, solar matter effects can significantly change the flavor conversions of solar neutrinos~\cite{Maltoni:2015kca}. However, for the low-energy $pp$ and $^7{\rm Be}$ neutrinos under consideration, the matter effects are insignificant. Even for the higher-energy ${\rm {}^7 Be}$ neutrinos, the solar matter effect on flavor conversions is as small as $4\%$, while the ${\rm {}^7 Be}$ neutrinos constitute only about $7\%$ of the total flux. For the $pp$ neutrinos, the matter effect will be much smaller, i.e., less than $1\%$. Although the matter effects are ignorable for solar $pp$ and ${\rm {}^7 Be}$ neutrinos, we include them in our numerical calculations for completeness.

\section{Constraints on Neutrino Lifetimes}

As the solar matter can hardly affect the low-energy neutrinos, after their production in the center of the Sun, they will loose quantum coherence during the long flight from the solar center to the solar surface. Therefore, it is safe to assume that low-energy neutrinos come out of the Sun as pure mass eigenstates. The fluxes of three neutrino mass eigenstates $\nu^{}_i$ (for $i = 1, 2, 3$) can be simply calculated by projecting the flux of the flavor eigenstate $\nu^{}_e$ onto the mass eigenstates. For the standard parametrization of the leptonic flavor mixing matrix $U$~\cite{Tanabashi:2018oca}, we have $|U^{}_{e1}|^2 = \cos^2 \theta^{}_{13} \cos^2 \theta^{}_{12}$, $|U^{}_{e2}|^2 = \cos^2 \theta^{}_{13} \sin^2 \theta^{}_{12}$ and $|U^{}_{e3}|^2 = \sin^2 \theta^{}_{13}$, where $\theta^{}_{13} \approx 8^\circ$ and $\theta^{}_{12} \approx 34^\circ$ are two of the three neutrino flavor mixing angles. Since the component of $\nu^{}_3$ is determined by the small number $\sin^2 \theta^{}_{13} \approx 0.02$, the impact of $\nu^{}_3$ decays on the constraints on the lifetimes of $\nu^{}_1$ and $\nu^{}_2$ can be neglected~\cite{Maltoni:2015kca}. For clarity, we take $\nu^{}_3$ to be absolutely stable and focus on the other two mass eigenstates $\nu^{}_1$ and $\nu^{}_2$ in this work. The effects of $\nu^{}_3$ decays will be briefly discussed in Section 3.3.

\subsection{Invisible Decays}

First, let us start with the IVD scenario, in which the decay products are completely invisible to the detector. In this case, the daughter particles could be sterile neutrinos or some other exotic particles in the hidden sector. As a consequence, the decaying neutrino mass eigenstates will simply disappear, and the survival probabilities of three neutrino mass eigenstates read
\begin{eqnarray} \label{eq:IVDprob}
P^{}_{e1} &=& \cos^2 \theta^{}_{13} \cos^2 \theta^{}_{12} \mathrm{exp} \left(-\frac{L}{E} \cdot \frac{m^{}_1}{\tau^{}_1} \right) \; , \nonumber \\
P^{}_{e2} &=& \cos^2 \theta^{}_{13} \sin^2 \theta^{}_{12} \mathrm{exp} \left(-\frac{L}{E} \cdot \frac{m^{}_2}{\tau^{}_2} \right) \; , \\
P^{}_{e3} &=& \sin^2 \theta^{}_{13} \; , \nonumber
\end{eqnarray}
where the exponential factors arise from neutrino decays, $L \approx 1.496 \times 10^8~{\rm km}$ is the distance from the Sun to the Earth~\footnote{As the distance from the Sun to the Earth is about $10^{3}$ times longer than the solar diameter, we can safely take this value as the distance from the production point of solar neutrinos to the detector at the Earth.}, $\tau^{}_i$ is the lifetime of $\nu^{}_i$ with the mass $m^{}_i$ (for $i = 1, 2$). To include the solar matter effect, we have to replace the mixing angles of Eq.~(\ref{eq:IVDprob}) in vacuum by their effective values in the solar matter, i.e., $\theta^{}_{ij} \rightarrow \tilde{\theta}^{}_{ij}$ (for $ij = 12, 13$). The effective mixing angles $ \tilde{\theta}^{}_{ij}$ can be solved by using the realistic density profile of the Sun. One can see Ref.~\cite{Khan:2017djo} for more discussions. Note that the IVDs of $\nu^{}_3$ have been ignored in Eq.~(\ref{eq:IVDprob}), as we have explained before. The differential fluxes of neutrino mass eigenstates at the Earth can be obtained via
 \begin{align}\label{eq:phimass}
\frac{{\rm d} \phi ^{}_{i}}{{\rm d} E} = \frac{\mathrm{d} \phi  }{\mathrm{d} E} \cdot P^{}_{ei} \; ,
 \end{align}
 for $i=1,2,3$, where ${\mathrm{d} \phi^{}_{} }/{\mathrm{d} E}$ is the original solar neutrino flux of $\nu^{}_{e}$, given by Eq.~(\ref{eq:Be7flux}) for the $^7{\rm Be}$ neutrinos and Eq.~(\ref{eq:ppflux}) for the $pp$ neutrinos. For the neutrino detection at the Earth, the mass eigenstates are projected onto their flavor states according to
  \begin{align}\label{eq:phiflavor}
\frac{{\rm d} \phi ^{}_{\alpha}}{{\rm d} E} = \sum^{}_{i} \frac{\mathrm{d} \phi^{}_{i} }{\mathrm{d} E} \cdot |U^{}_{\alpha i}|^2 \; ,
 \end{align}
for $\alpha=e,\mu,\tau$. The fluxes in Eq.~(\ref{eq:phiflavor}) are then used in Eq.~(\ref{eq:events}) to calculate the ER event rate.

\begin{figure}[t!]
    \begin{center}
    \hspace{-0.2cm}
    \subfigure{%
    \includegraphics[width=0.47\textwidth]{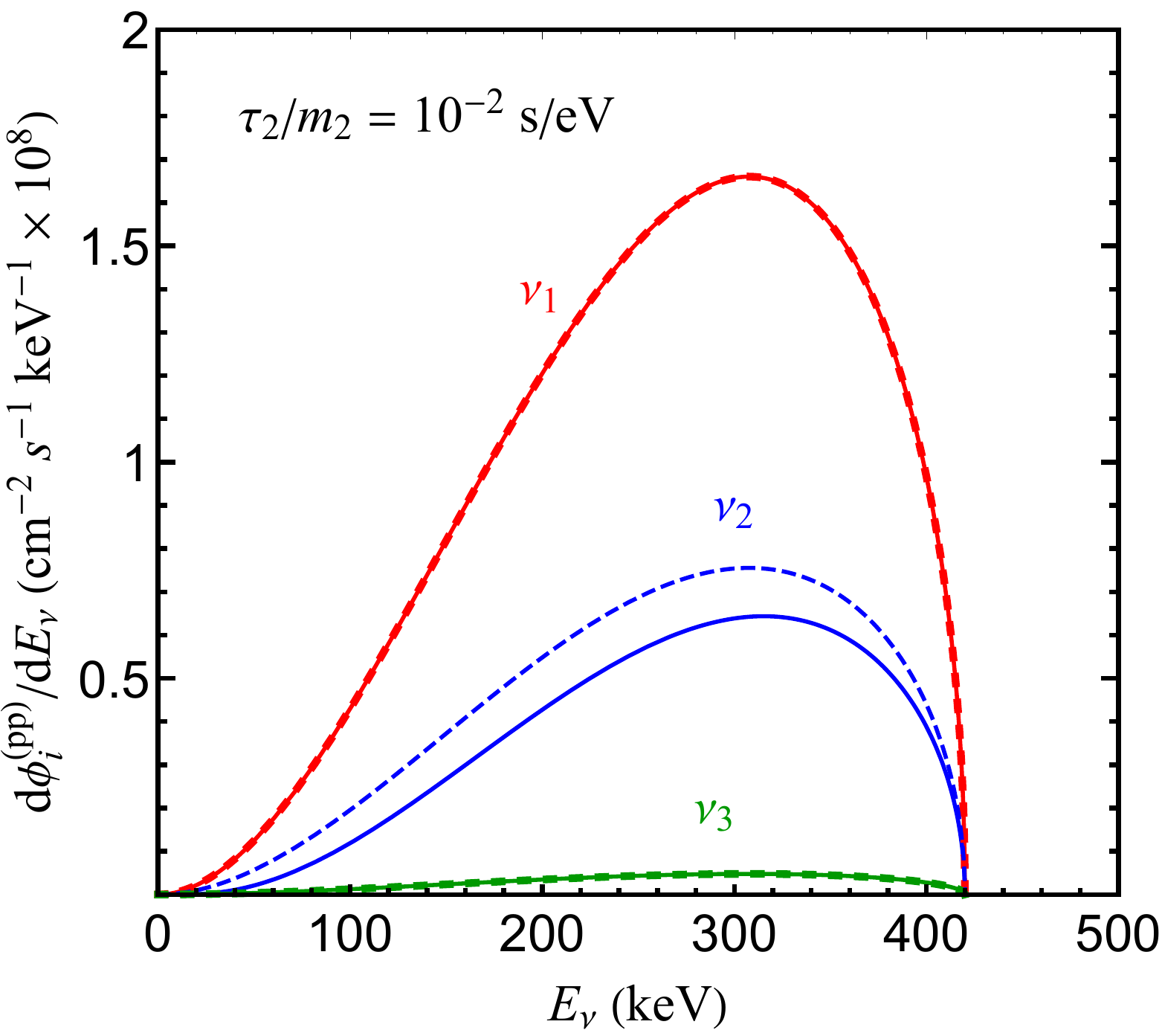}
    }%
    \subfigure{%
    \hspace{0.5cm}
    \includegraphics[width=0.445\textwidth]{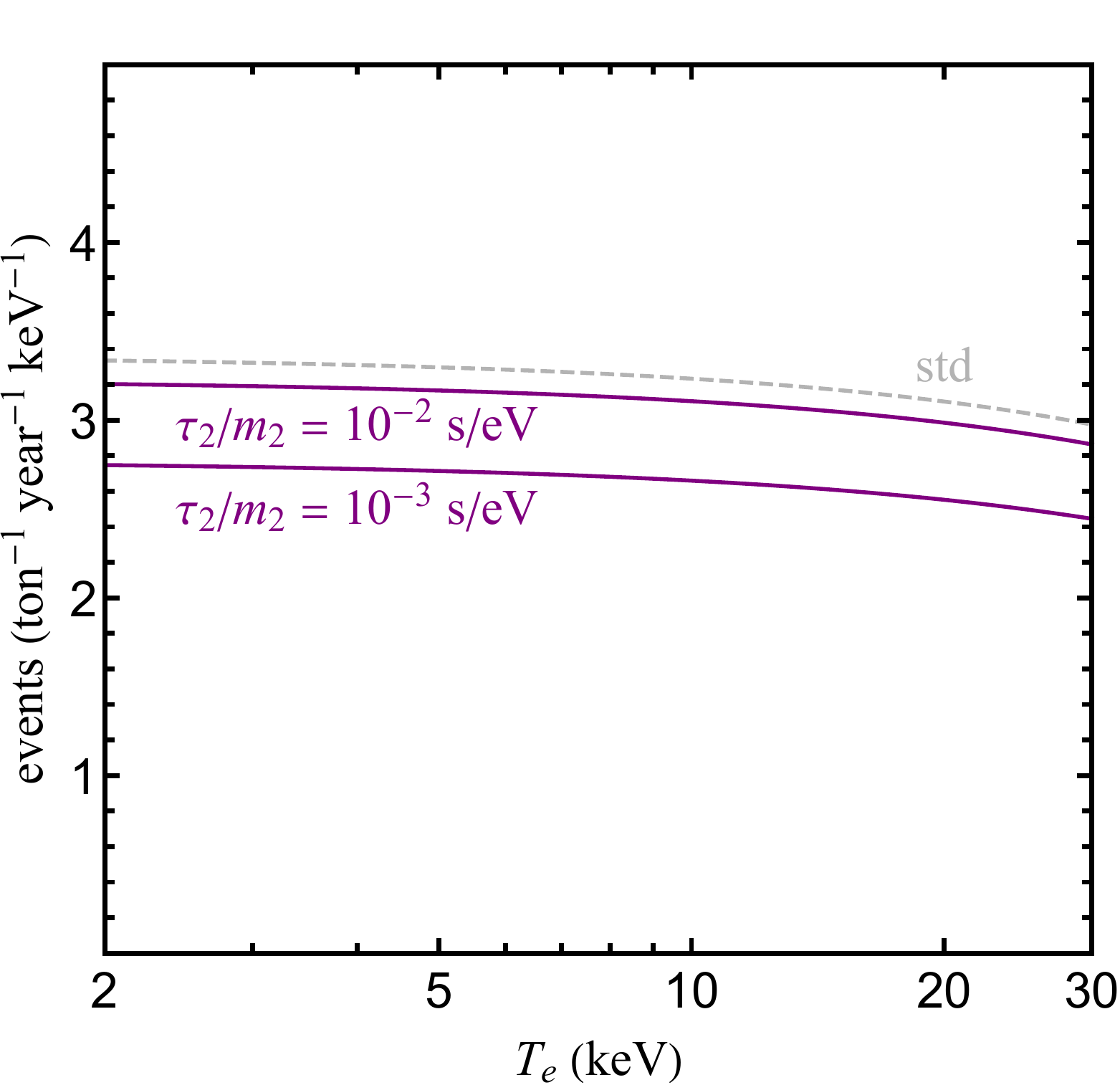}        }
    \end{center}
    \vspace{-0.5cm}
    \caption{The energy spectra of neutrino mass eigenstates $\nu^{}_1$ (red curves), $\nu^{}_2$ (blue curves) and $\nu^{}_3$ (green curves) arriving at the Earth from solar $pp$ neutrinos are presented in the case of invisible decays with $\tau^{}_2/m^{}_2 = 10^{-2}~{\rm s}/{\rm eV}$ ({\it Left Panel}). The dashed curves stand for the case of stable neutrinos and the solid ones for the case of neutrino decays. The event spectra of the elastic neutrino-electron scattering in the xenon detector are calculated for the cases with $\tau^{}_2/m^{}_2 = 10^{-2}~{\rm s/eV}$ and $\tau^{}_2/m^{}_2 = 10^{-3}~{\rm s/eV}$, where the latter corresponds to the case of a nearly complete decay of $\nu^{}_1$ ({\it Right Panel}).}
    \label{fig:fig1}
    \end{figure}

In Fig.~1, we have shown the energy spectra of $\nu^{}_i$ from the solar $pp$ neutrinos (the left panel) and the event spectra of the elastic neutrino-electron scattering in the DARWIN detector in the energy window of $(2 \cdots 30)~{\rm keV}$ (the right panel). In both panels of Fig.~\ref{fig:fig1}, the dashed curves stand for the standard case of stable neutrinos, while the solid curves for the cases of neutrino decays. For illustration, only the IVDs of $\nu^{}_1$ are considered for $\tau^{}_2/m^{}_2 = 10^{-2}~{\rm s/eV}$ or $\tau^{}_2/m^{}_2 = 10^{-3}~{\rm s/eV}$. Since the IVDs of neutrino mass eigenstates lead to a less number of active neutrinos at the detector, the number of ER events should be reduced accordingly, as clearly indicated in the right panel of Fig.~1.

\subsection{Non-Radiative Visible Decays}

Then, we proceed to investigate the case of NRVDs, for which the energy spectra of neutrino mass eigenstates will be changed in a relatively more complicated way. Such kinds of decays take place in various new-physics models, and can be represented by $\nu^{}_i \to \nu^{}_j + \eta$, where $\eta$ denotes the particle invisible in the detector. In the majoron models, $\eta$ could be a massless or very light pseudo-scalar particle. However, our results will also be applicable to a more general case with $\eta$ being either a scalar or vector particle. The NRVDs of neutrinos are more interesting in several aspects, such as the roles played by the unknown neutrino mass ordering and neutrino mass spectrum, which have not been sufficiently emphasized in the previous works~\cite{Beacom:2002cb, Joshipura:2002fb, Bandyopadhyay:2002qg, Berryman:2014qha, Picoreti:2015ika}. Therefore, we pay more attention to this scenario and carry out a more careful study.

\subsubsection{Normal mass ordering}

If the normal neutrino mass ordering with $m^{}_1 < m^{}_2 < m^{}_3$ (NO) is assumed, the lightest neutrino mass eigenstate $\nu^{}_1$ will be stable. Then, we have two different decay channels for the heaviest state $\nu^{}_3$, i.e., $\nu^{}_3 \to \nu^{}_1 + \eta$ and $\nu^{}_3 \to \nu^{}_2 + \eta$, and only one decay channel for $\nu^{}_2$, i.e., $\nu^{}_2 \to \nu^{}_1 + \eta$. Since the component of $\nu^{}_3$ is highly suppressed in solar neutrinos, the only relevant decay channel is just $\nu^{}_2 \to \nu^{}_1 + \eta$. As we shall explain below, the energy of the final-state $\nu^{}_1$ in the laboratory frame depends on the mass spectrum of $\nu^{}_1$ and $\nu^{}_2$~\cite{Beacom:2002cb}, given a vanishing mass of $\eta$ (as in a class of majoron models).

In the rest frame of $\nu^{}_2$, the energies of $\eta$ and $\nu^{}_1$ are given by $E^\prime_\eta = |{\bf p}^\prime|= (m^2_2 - m^2_1)/(2 m^{}_2)$ and $E^\prime_1 = \sqrt{|{\bf p}^\prime|^2 + m^2_1}$ with ${\bf p}^\prime$ being the three-momentum of $\eta$, respectively. In the laboratory frame, the energy of $\eta$ turns out to be $E^{}_\eta = \gamma (E^\prime_\eta + \beta |{\bf p}^\prime| \cos \theta)$, where the Lorentz factor is $\gamma = E^{}_2/m^{}_2$ and $\beta = |{\bf p}^{}_2|/E^{}_2$ with ${\bf p}^{}_2$ being the three-momentum of the decaying neutrino $\nu^{}_2$. Here $\theta$ is the angle between ${\bf p}^\prime$ and ${\bf p}^{}_2$. Note that neutrino masses are much smaller than neutrino energies in the laboratory frame, i.e., $m^{}_i \ll E^{}_i$, which is not far from reality in the case of solar neutrinos. Therefore, it is straightforward to figure out
\begin{eqnarray} \label{eq:eeta}
E^{}_\eta = \frac{E^{}_2}{2} \left(1 - \frac{m^2_1}{m^2_2}\right) \left(1 + \sqrt{1 - \frac{m^2_2}{E^2_2}} \cos\theta \right) \approx \frac{E^{}_2}{2} \left(1 - \frac{m^2_1}{m^2_2}\right) \left(1 + \cos\theta\right) \; ,
\end{eqnarray}
from which one can obtain the minimum and maximum energies of $\eta$, namely, $E^{\rm min}_\eta \approx 0$ and $E^{\rm max}_\eta \approx E^{}_2 (1 - m^2_1/m^2_2)$. Because of the energy conservation $E^{}_2 = E^{}_1 + E^{}_\eta$, we can observe that the energy of $\nu^{}_1$ should be in the range of $(m^2_1/m^2_2) E^{}_2 \leq E^{}_1 \leq E^{}_2$, and similar expressions can be found in Ref.~\cite{Balantekin:2018ukw}. As an immediate consequence, the daughter neutrino $\nu^{}_1$ will carry away most of the energy of $\nu^{}_2$ when the mass ratio $m^{}_1/m^{}_2$ is close to one~\cite{Beacom:2002cb}.

\begin{figure}[t!]
    \begin{center}
    \hspace{-0.2cm}
    \subfigure{%
    \includegraphics[width=0.47\textwidth]{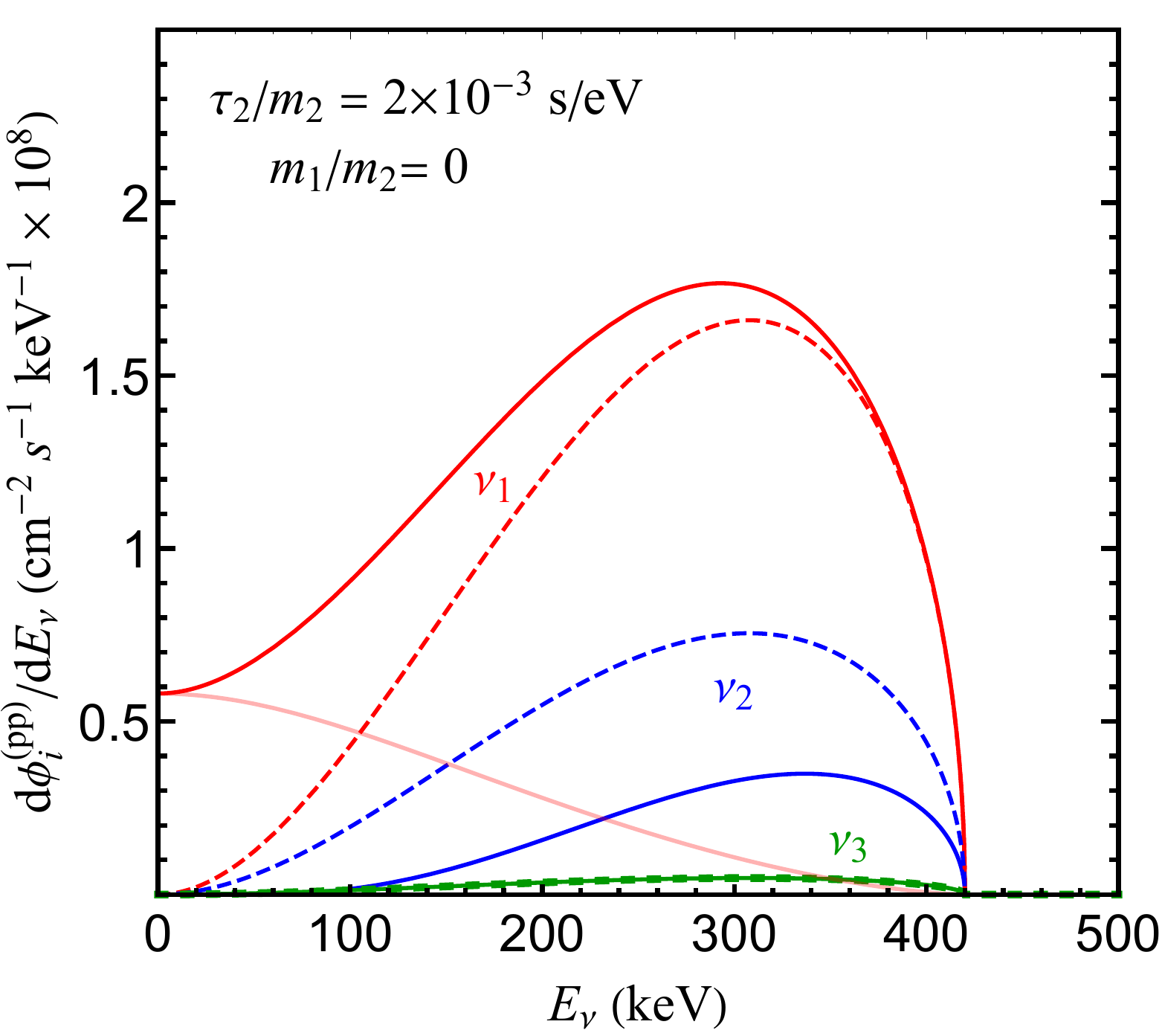}
    }%
    \subfigure{%
    \hspace{0.2cm}
    \includegraphics[width=0.47\textwidth]{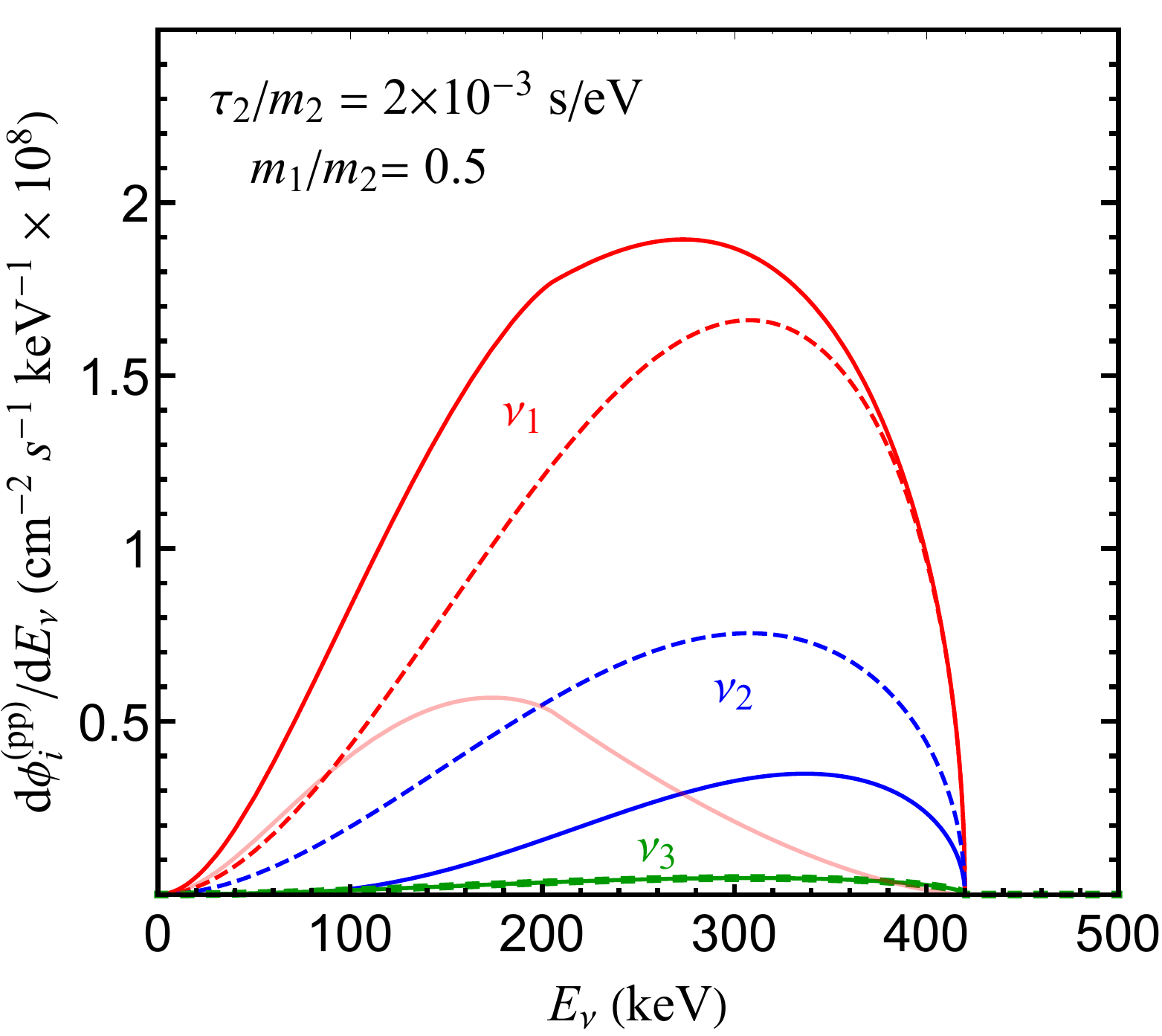}        }
    \subfigure{%
    \includegraphics[width=0.47\textwidth]{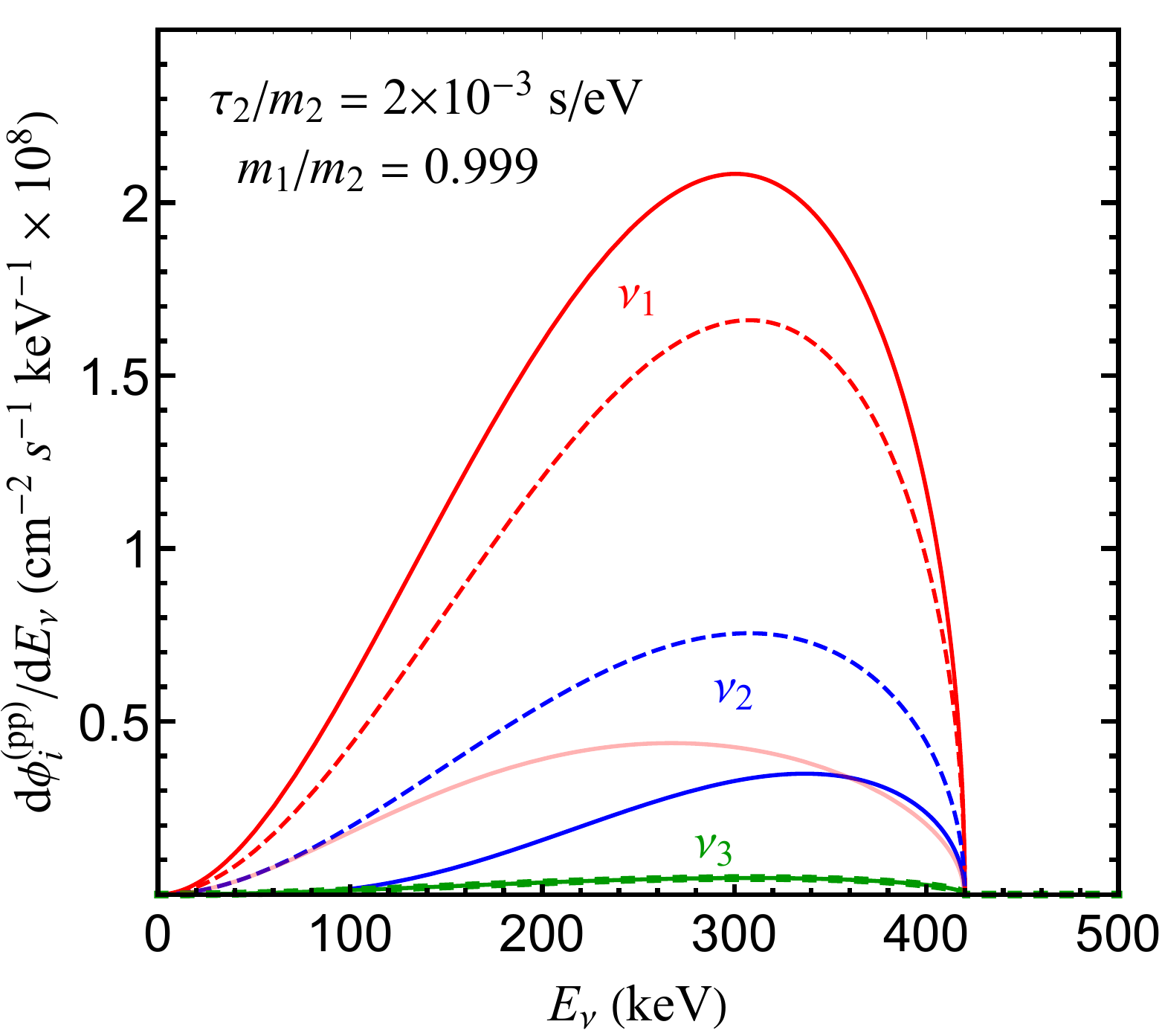}
    }%
    \subfigure{%
    \hspace{0.2cm}
    \includegraphics[width=0.46\textwidth]{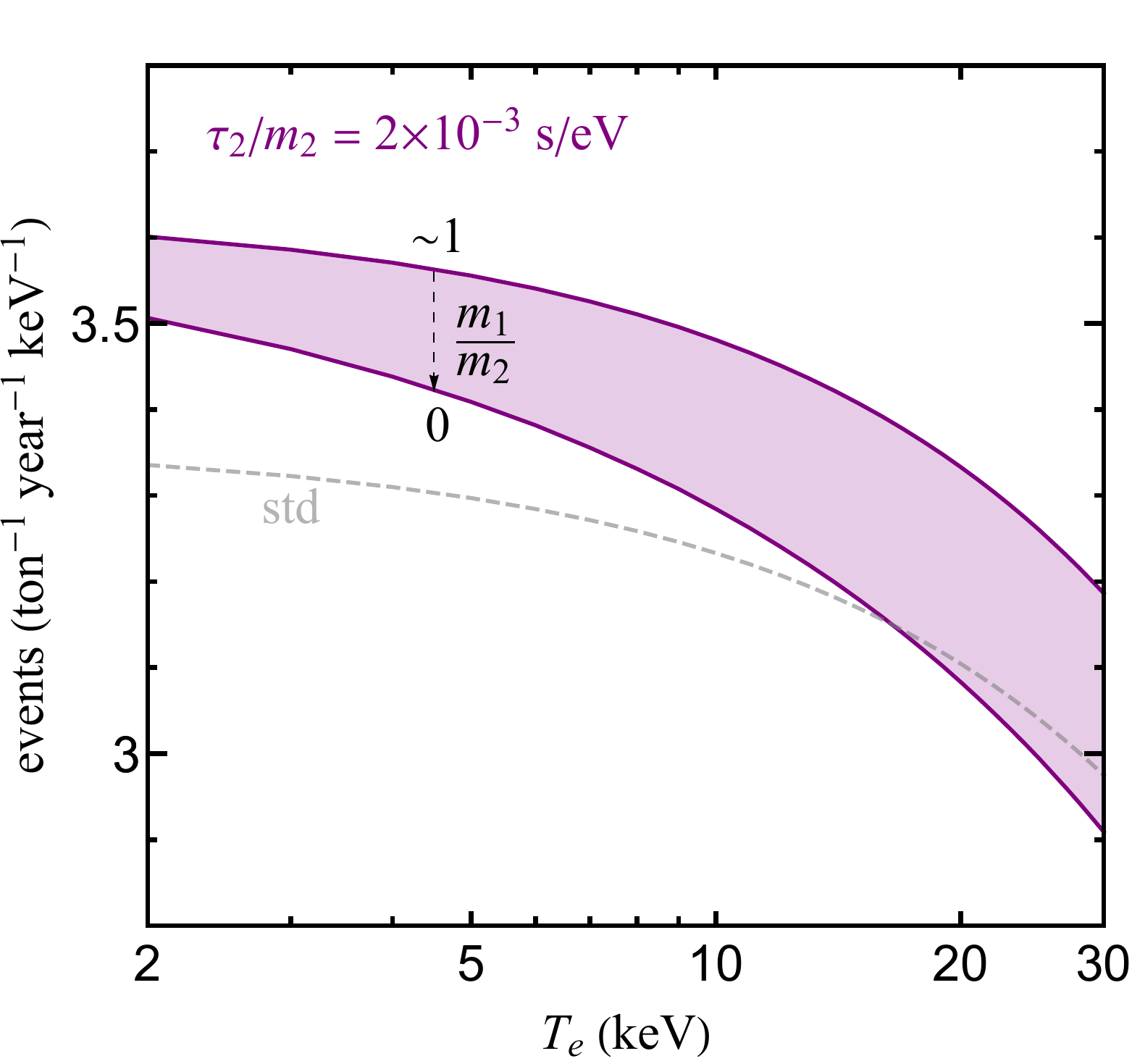}        }
    \end{center}
    \vspace{-0.5cm}
    \caption{The energy spectra of neutrino mass eigenstates arriving at the Earth from solar $pp$ neutrinos in the case of non-radiative visible decays with normal neutrino mass ordering but different mass ratios: (1) $m^{}_1/m^{}_2 = 0$ ({\it Upper-left Panel}), (2) $m^{}_1/m^{}_2 = 0.5$ ({\it Upper-right Panel}), and (3) $m^{}_1/m^{}_2 = 0.999$ ({\it Lower-left Panel}), where $\nu^{}_2 \to \nu^{}_1 + \eta$ with $\tau^{}_2/m^{}_2 = 2\times 10^{-3}~{\rm s}/{\rm eV}$ is assumed. The dashed curves stand for the case of stable neutrinos and the solid ones for the case of neutrino decays. The light red curves represent the contribution to the $\nu^{}_1$ spectrum from the decays of $\nu^{}_2$'s. The event spectra for different values of $m^{}_1/m^{}_2 \in [0, 1)$ have been given in the lower-right panel.}
    \label{fig:fig2}
\end{figure}
The energy distribution of the produced $\nu^{}_1$ in the laboratory frame is determined by the angular distribution of $\nu^{}_1$ in the rest frame of $\nu^{}_{2}$, while the angular distribution depends on the properties of the interaction between neutrinos and the invisible particle $\eta$. Given the interaction forms, one can immediately find the expressions of the energy dependence of the decay rate in Refs.~\cite{Coloma:2017zpg,Kim:1990km,Berezhiani:1991vk,Giunti:1992sy,Lindner:2001fx}. Without specifying any new-physics models, we assume that the angular distribution of final-state particles is isotropic in the rest frame of $\nu^{}_2$. It is very straightforward to extend our discussions to any specific models. However, as our main purpose in this work is to study the general sensitivities of the next generation xenon detectors, we will not elaborate on a specific model of neutrino decays. Therefore, in the assumption of isotropic angular distribution, the decay rate of $\nu^{}_{2}$ has no dependence on the energy of the final $\nu^{}_{1}$ according to Eq.~(\ref{eq:eeta}). After taking account of the decays $\nu^{}_2 \to \nu^{}_1 + \eta$ and the resultant energy distribution of $\nu^{}_1$, we can find out the energy spectra of neutrino mass eigenstates at the Earth
\begin{eqnarray}
\frac{\mathrm{d} \phi^{}_2}{\mathrm{d} E} &=& \frac{\mathrm{d} \phi^0_2}{\mathrm{d} E} \mathrm{exp} \left({-\frac{L}{E}\cdot \frac{m^{}_2}{\tau^{}_{2}}}\right) \; , \nonumber \\
\frac{\mathrm{d} \phi^{}_1}{\mathrm{d} E} &=& \frac{\mathrm{d} \phi^0_1}{\mathrm{d} E} + \int_{E^\prime_{\rm min}}^{E^\prime_{\rm max}} \frac{\mathrm{d} \phi^0_2}{\mathrm{d} E^\prime} \left[1 - \mathrm{exp}\left({-\frac{L}{E^\prime} \cdot \frac{m^{}_2}{\tau^{}_{2} }}\right)\right] \frac{1}{1 - m^2_1/m^2_2} \frac{\mathrm{d}E^\prime}{E^\prime} \; ,
\end{eqnarray}
where ${\rm d} \phi^0_i/{\rm d} E = {\mathrm{d} \phi}/{\mathrm{d} E} \cdot |U^{}_{ei}|^2 $ (for $i = 1, 2$) stand for the reference spectra without neutrino decays, $E^\prime_{\rm min} = E$, and $E^\prime_{\rm max} = (m_2^2/m_1^2)E$. In Fig.~\ref{fig:fig2}, the energy spectra of $\nu^{}_i$ (for $i = 1, 2, 3$) are displayed for different values of the mass ratio $m^{}_1/m^{}_2$: (1) the hierarchical scenario with $m^{}_1/m^{}_2 = 0$ (the upper-left panel); (2) the intermediate scenario with $m^{}_1/m^{}_2 = 0.5$ (the upper-right panel); and (3) the nearly-degenerate scenario with $m^{}_1/m^{}_2 = 0.999$ (the lower-left panel). Some comments on the numerical results are in order:
\begin{itemize}
\item In the hierarchical scenario with $m^{}_1/m^{}_2 = 0$, one can observe from the upper-left panel of Fig.~\ref{fig:fig2} that the low-energy end of the spectrum of $\nu^{}_1$ is enhanced significantly. The main reason is that $\nu^{}_2$'s with much lower energies are more likely to decay for a given lifetime $\tau^{}_2/m^{}_2 = 2\times 10^{-3}~{\rm s}/{\rm eV}$ during the flight to the Earth. The contribution to the $\nu^{}_1$ spectrum from the decays of $\nu^{}_2$'s has been represented by the light red curve. In other two scenarios, similar curves have also been drawn for reference.

\item For $m^{}_1/m^{}_2 = 0.5$, although the lower-energy $\nu^{}_2$'s will decay into $\nu^{}_1$'s, the latter will be more energetic as we have mentioned before. Therefore, the final $\nu^{}_1$ spectrum has been shifted towards the high-energy part. On the other hand, the cross section of the elastic neutrino-electron scattering increases for higher neutrino energies, so the event rate turns out to be larger than that in the previous case of a smaller mass ratio.

\item For the nearly-degenerate scenario with $m^{}_1/m^{}_2 = 0.999$, the final energy spectrum of $\nu^{}_1$'s is further modified via the contribution from the decays of $\nu^{}_2$'s and the event rate becomes even larger. The event rates for an arbitrary mass ratio in the whole range of $m^{}_1/m^{}_2 \in [0, 1)$ have been given in the lower-right panel of Fig.~\ref{fig:fig2}. However, it should be noticed that the lifetime $\tau^{}_2/m^{}_2 = 2\times 10^{-3}~{\rm s}/{\rm eV}$ is always fixed. In the limit of $m^{}_1 \to m^{}_2$, the phase space for the decay process $\nu^{}_2 \to \nu^{}_1 + \eta$ becomes vanishingly small, one has to increase the coupling between neutrinos and $\eta$ in order to guarantee a sizable decay rate $\Gamma^{}_2 \equiv \tau^{-1}_2$. In a specific model, we could run into the problem that an extremely large coupling constant leads to the invalidity of a perturbative calculation of the decay rate. Consequently, although the event rate in the nearly-degenerate scenario is much larger than that in the standard case without neutrino decays, we shall not expect a severe constraint on the coupling constant.
\end{itemize}
If a shorter lifetime $\tau^{}_2/m^{}_2 = 2\times 10^{-4}~{\rm s}/{\rm eV}$ is considered, $\nu^{}_2$'s could almost completely decay into $\nu^{}_1$'s and we find that the deviation of the event rate from that in the standard case can reach about $6\%$ in the optimal scenario. For more general cases, we leave a thorough study of the experimental sensitivities to neutrino lifetimes for Section 3.3.

\subsubsection{Inverted mass ordering}

Now we come to the inverted neutrino mass ordering (IO) with $m^{}_3 < m^{}_1 < m^{}_2$. In this case, the possible decay channels are (1) $\nu^{}_2 \to \nu^{}_1 + \eta$ and $\nu^{}_2 \to \nu^{}_3 + \eta$, with the partial decay widths $\Gamma^{}_{21} \equiv \tau^{-1}_{21}$ and $\Gamma^{}_{23} \equiv \tau^{-1}_{23}$, respectively; and (2) $\nu^{}_1 \to \nu^{}_3 + \eta$, with the decay width $\Gamma^{}_1 \equiv \tau^{-1}_1$. Since the cross section of $\nu^{}_3$'s interacting with electrons in the detector is suppressed by $\sin^2 \theta^{}_{13} \approx 0.02$, the decays of $\nu^{}_2$ and $\nu^{}_1$ into $\nu^{}_3$ are similar to the IVDs. In addition, given $\Delta m^2_{21} \equiv m^2_2 - m^2_1 \approx 7.5\times 10^{-5}~{\rm eV}^2$ and $\Delta m^2_{31} \equiv m^2_3 - m^2_1 \approx -2.5\times 10^{-3}~{\rm eV}^2$ in the IO case, one can calculate the neutrino mass ratio
\begin{eqnarray}
\frac{m^{}_1}{m^{}_2} = \sqrt{\frac{m^2_3 + |\Delta m^2_{31}|}{m^2_3 + \Delta m^2_{21} + |\Delta m^2_{31}|}} \gtrsim 0.985 \; ,
\end{eqnarray}
implying a nearly-degenerate mass spectrum for $\nu^{}_1$ and $\nu^{}_2$. As we have discussed in the previous section, $\nu^{}_1$ will take away most of the energy of $\nu^{}_2$ in the decay of $\nu^{}_2 \to \nu^{}_1 + \eta$ with $\eta$ being a massless particle. In consideration of $\nu^{}_1$ and $\nu^{}_2$ decaying into $\nu^{}_3$, the mass ratios $m^{}_3/m^{}_2 \approx m^{}_3/m^{}_1$ are important to derive the final energy spectrum of $\nu^{}_3$'s. As we have pointed out, however, the interaction of the latter with electrons is highly suppressed and we need not to worry about the true neutrino mass spectrum.

Without loss of generality,
\begin{figure}[t!]
    \begin{center}
    \includegraphics[width=0.47\textwidth]{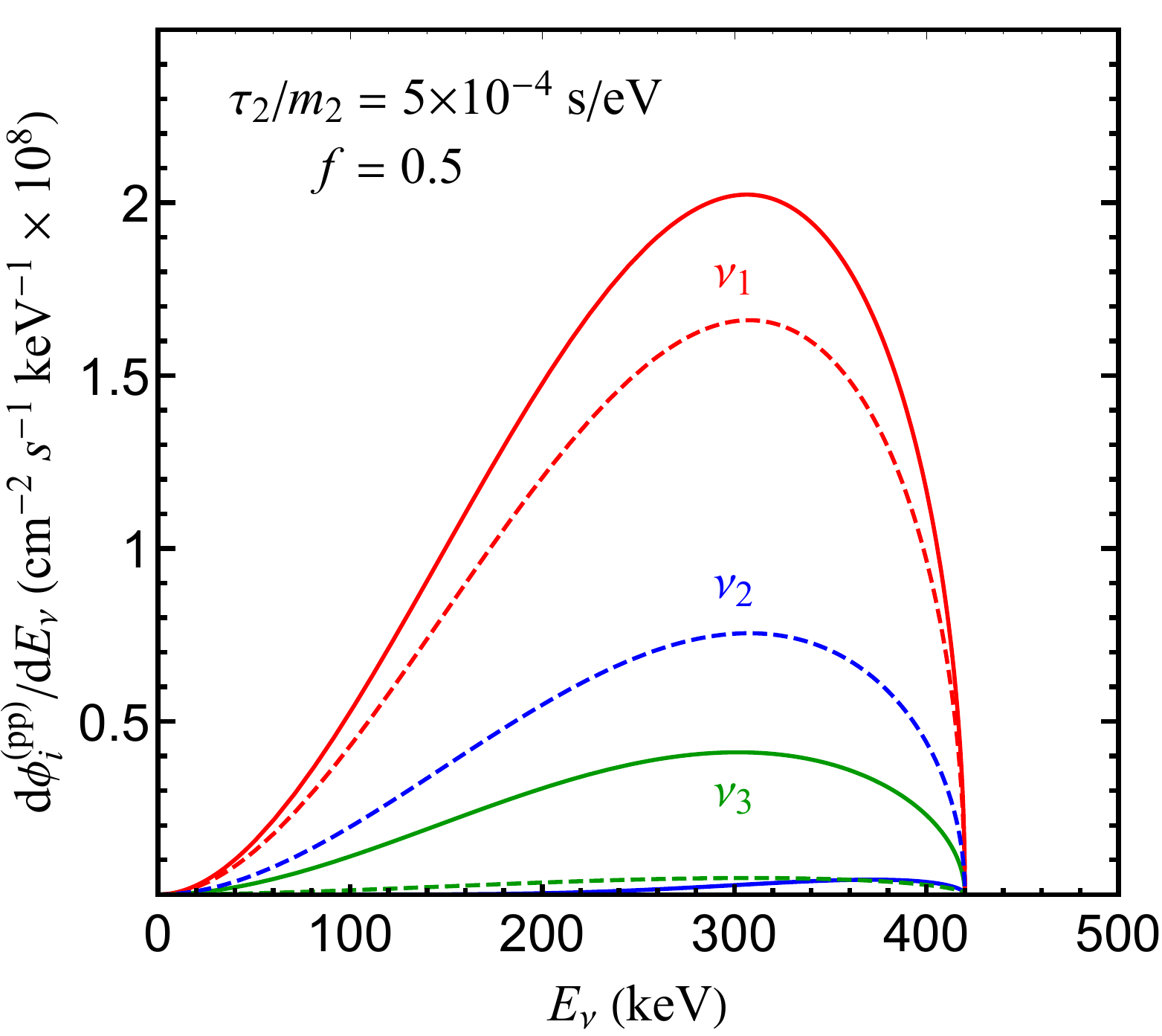}
    \hspace{0.5cm}
    \includegraphics[width=0.45\textwidth]{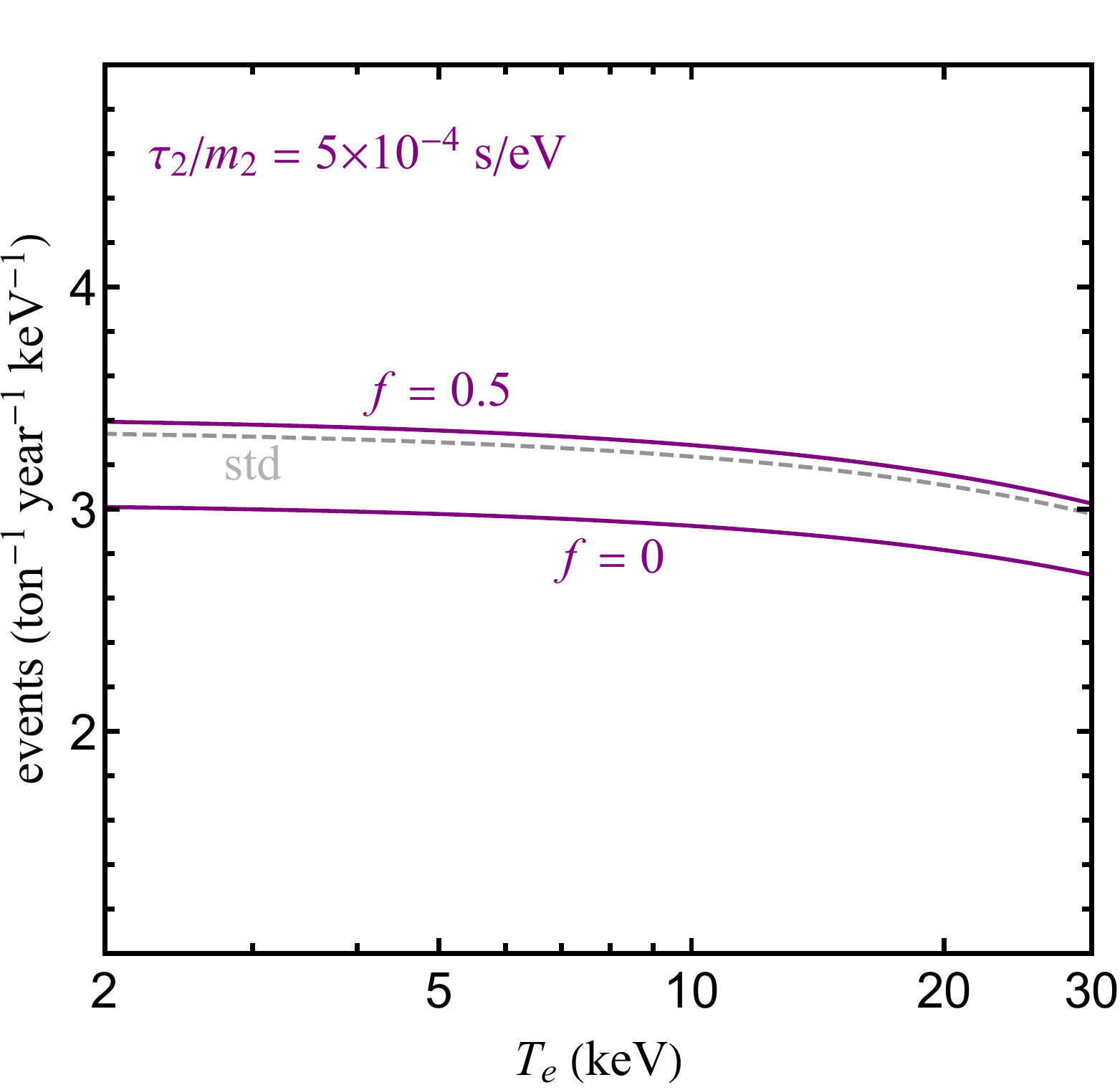}
    \end{center}
    \vspace{-0.5cm}
    \caption{The energy spectra of neutrino mass eigenstates arriving at the Earth from solar $pp$ neutrinos in the case of non-radiative visible decays with inverted neutrino mass ordering, where the branching fraction $f = 0.5$ for $\nu^{}_2 \to \nu^{}_1 + \eta$ and $\tau^{}_2/m^{}_2 = 5\times 10^{-4}~{\rm s}/{\rm eV}$ are assumed ({\it Left Panel}). The dashed curves stand for the case of stable neutrinos and the solid ones for the case of neutrino decays. The event spectra in the cases of $f= 0$ and $f = 0.5$ have been calculated for $\tau^{}_2/m^{}_2 = 5\times 10^{-4}~{\rm s}/{\rm eV}$ ({\it Right Panel}).}
    \label{fig:fig3}
\end{figure}
we take the neutrino mass spectrum to be nearly degenerate (i.e., $m^{}_3 \sim m^{}_1 \sim m^{}_2$). The survival probabilities of three neutrino mass eigenstates are found to be
\begin{eqnarray}\label{eq:IOprob}
P^{}_{e2} &=& \cos^2 \theta^{}_{13} \sin^2 \theta^{}_{12} \mathrm{exp}\left(-\frac{L}{E} \cdot \frac{m^{}_{2}}{\tau^{}_{2}} \right) \; , \nonumber \\
P^{}_{e1} &=& \cos^2 \theta^{}_{13} \cos^2 \theta^{}_{12} \mathrm{exp}\left(-\frac{L}{E} \cdot \frac{m^{}_{1}}{\tau^{}_{1}} \right) + \frac{P^{}_{e2} f}{\left(\displaystyle 1 - \frac{m^{}_1}{\tau^{}_1} \cdot \frac{\tau^{}_2}{m^{}_2}\right)} \left\{ {\rm exp}\left[\left(\frac{m^{}_2}{\tau^{}_2} - \frac{m^{}_1}{\tau^{}_1}\right)\frac{L}{E}\right] - 1 \right\} \; ,~~~~~~~ ~~~~~~~~~~~ \quad
\end{eqnarray}
and $P^{}_{e3} = 1 - P^{}_{e2} - P^{}_{e1}$ \footnote{Since the neutrino mass spectrum is assumed to be nearly degenerate in the IO case, the daughter neutrino will have an energy almost the same as the parent neutrino energy. For the scenario of NRVDs, the total survival probability of three active neutrinos at any given energy is conserved, e.g., the number of the decayed $\nu^{}_{1}$ via the process $\nu^{}_1 \to \nu^{}_3 + \eta$ will be fully converted to that of the $\nu^{}_{3}$ being produced. Therefore, it is a good approximation to have the relation $P^{}_{e1} + P^{}_{e2} + P^{}_{e3} = 1$, as can also be seen from Eq.~(\ref{eq:probconser}) in Appendix A. The situation is different from Ref.~\cite{Berryman:2014yoa} where the active neutrinos are assumed to decay into purely invisible components as in our IVD scenario.}, where $f \equiv \Gamma^{}_{21}/(\Gamma^{}_{21} + \Gamma^{}_{23})$ is the branching ratio of the $\nu^{}_2 \rightarrow \nu^{}_1$ decay. The derivation of Eq.~(\ref{eq:IOprob}) can be found in Appendix A. As has been mentioned before, the energy spectra ${\rm d}\phi^{}_i/{\rm d}E$ of neutrino mass eigenstates can be obtained via ${\rm d}\phi/{\rm d}E \cdot P^{}_{ei}$ for $i = 1, 2, 3$, where ${\rm d}\phi/{\rm d}E$ denotes the original spectrum of solar neutrinos $\nu^{}_e$. In Eq.~(\ref{eq:IOprob}), the second term in the expression for $P^{}_{e1}$ arises from the interplay between the production of $\nu^{}_1$ from the $\nu^{}_2$ decays and the subsequent decay of $\nu^{}_1$ itself.

In Fig.~\ref{fig:fig3},
we have given the energy spectra of neutrino mass eigenstates arriving at the Earth in the NRVD case with the IO (the left panel), and the corresponding event spectra have been calculated for two different branching ratios $f = 0$ and $f = 0.5$ (the right panel). In both panels, the lifetime $\tau^{}_2/m^{}_2 = 5\times 10^{-4}~{\rm s}/{\rm eV}$ has been assumed. Since neutrino oscillation data point to $|U^{}_{e1}|^2 \approx 0.67 > |U^{}_{e2}|^2 \approx 0.31$, which means that the interaction of $\nu^{}_1$ with electrons will be twice stronger than that of $\nu^{}_2$, the increase of the $\nu^{}_1$ number from the $\nu^{}_2\rightarrow \nu^{}_1$ decays will enhance the event rate. On the other hand, the $\nu^{}_2 \rightarrow \nu^{}_3$ and $\nu^{}_1 \rightarrow \nu^{}_3$ decays are almost equivalent to the invisible decays, reducing the event rate. As a result, for the branching fraction $f = 0.5$, both $\nu^{}_2 \to \nu^{}_1$ and $\nu^{}_{2} \to \nu^{}_3$ decays take place and the final event rate deviates only slightly from the standard one. The results for $f = 0$ are similar to those in the case of IVDs, as shown in the right panel of Fig.~\ref{fig:fig1}.

\subsection{Experimental Sensitivities}

After summarizing the main features of both IVDs and NRVDs in the previous sections, we now perform a statistical study of the experimental sensitivities to neutrino lifetimes in these scenarios. Taking the liquid xenon part of the DARWIN experiment for example, we assume a total liquid xenon mass of 21.4 tons, with a fiducial mass of 14 tons, and the ER energy window of $(2 \cdots 30)~{\rm keV}$. It has been found in Ref.~\cite{Baudis:2013qla} that such a fiducial mass of liquid xenon would be almost optimal to achieve a relatively large number of solar neutrino events while keeping a low background from the cryostat and detector materials. See Fig.~2 in Ref.~\cite{Baudis:2013qla} for more details. The major backgrounds, which have already been summarized in Table.~1 of Ref.~\cite{Schumann:2015cpa} and Table.~2 of Ref.~\cite{Baudis:2013qla}, are coming from the radioactivities of $^{85}{\rm Kr}$ and $^{222}{\rm Rn}$, the $\gamma$-ray materials, and the double-beta decays of $^{136}{\rm Xe}$. The corresponding background rates in the unit of counts per ${\rm ton}\cdot {\rm year} \cdot {\rm keV}$ are $1.44$ for $^{85}{\rm Kr}$, $0.35$ for $^{222}{\rm Rn}$, $0.054$ for $\gamma$-ray materials, and $0.12~(T^{}_e/1~{\rm keV})$ for double-beta decays of $^{136}{\rm Xe}$, which will be used in our sensitivity study. Since the ER events in the xenon detector cannot be fully discriminated from the NR events yet, the NR events from the neutrino floor should also be a background. However, its contribution to the background at the level of $\mathcal{O}(10^{-3})$ is negligible assuming no ER/NR discrimination~\cite{Baudis:2013qla}. This type of background can be further reduced by using the discrimination technique.
\begin{figure}[t!]
    \begin{center}
    \includegraphics[width=0.5\textwidth]{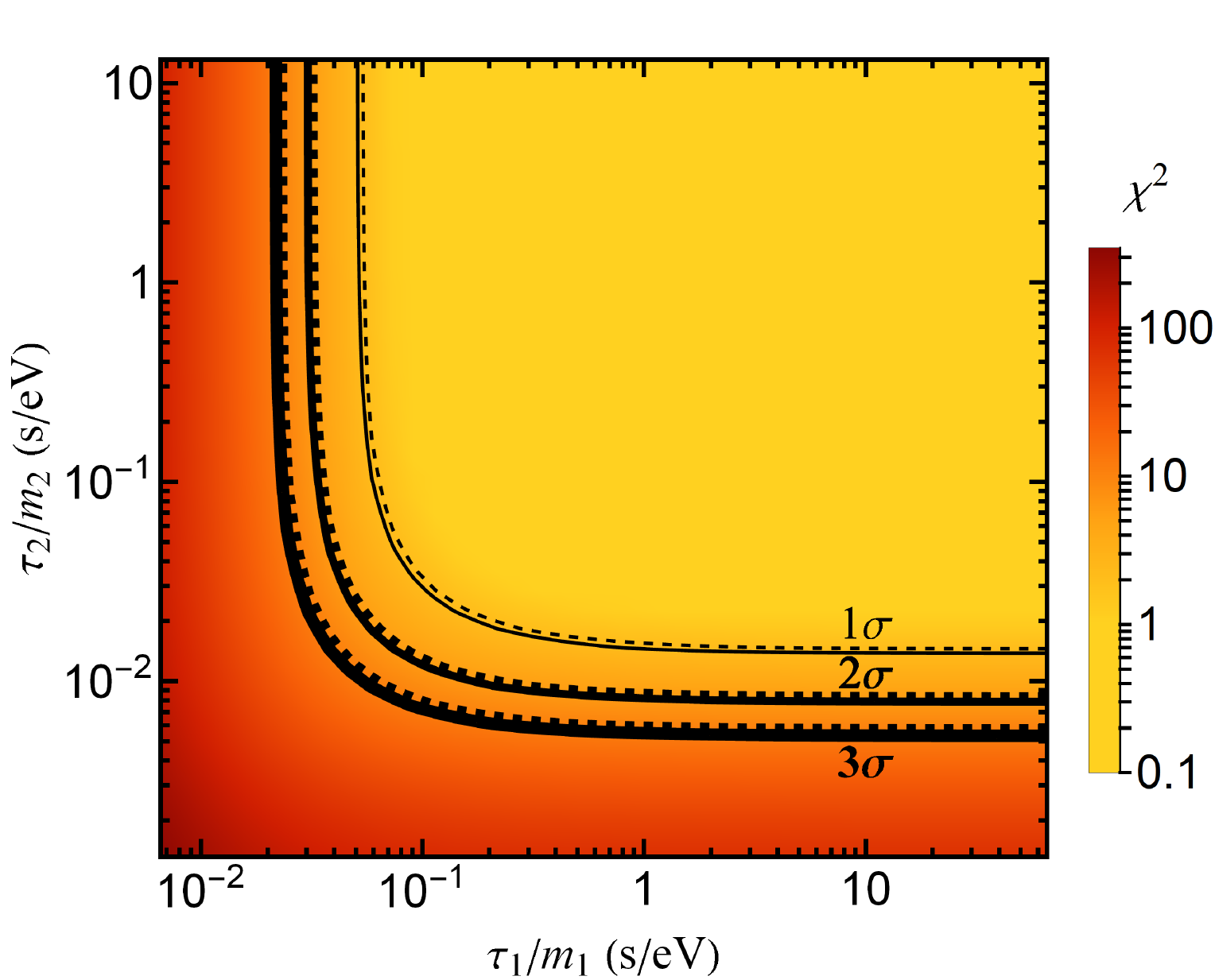}
    \end{center}
    \vspace{-0.5cm}
    \caption{The expected sensitivities of a DARWIN-like xenon experiment to neutrino lifetimes $(\tau^{}_1/m^{}_1, \tau^{}_2/m^{}_2)$ in the case of invisible decays, where the exposure of $70~{\rm ton} \cdot {\rm year}$ and the ER energy window $(2 \cdots 30)~{\rm keV}$ have been adopted. The solid (dashed) contours stand for the constraints with (without) considering theoretical errors of solar neutrino fluxes and $\theta^{}_{12}$, where the $1 \sigma$, $2\sigma$, and $3 \sigma$ contours are represented by thin, medium, and thick curves, respectively.}
    \label{fig:sen1}
\end{figure}

For the total exposure of $70~{\rm ton} \cdot {\rm year}$, equivalent to the detector with a fiducial mass of 14 tons running for five years, the expected number of solar neutrino events is about 6600, while that of the background events is about 7500, in the chosen energy window of $(2\cdots 30)~{\rm keV}$. The precision for the solar $pp$ neutrino flux can reach the level of $1.8\%$, which is about three times worse than current uncertainties in the theoretical predictions of the $pp$ neutrino flux. To properly take into account the distortion of the event spectrum in the hypothesis of neutrino decays, we divide the energy window of $(2 \cdots 30)~{\rm keV}$ into 28 bins in a uniform way. The theoretical prediction for the event number in the $i$th bin, i.e., $R^i(\tau/m)$, is calculated by assuming neutrino decays in each scenario, while the experimental observation $R^i_{\rm ex}$ is determined as in the standard case without neutrino decays. In order to characterize the experimental power to constrain neutrino decays, we construct the $\chi^2$ function as follows
\begin{align}
\chi^2 = \sum_{i,j}^{\rm bins} \left[R^i(\tau/m) - R^i_{\rm ex}\right] (\sigma^2)^{-1}_{ij} \left[R^j(\tau/m) - R^j_{\rm ex}\right] \; ,
\end{align}
where $\sigma^2$ is the covariance matrix of total errors. The statistical error $u^2_i$ for each energy bin is assumed to be uncorrelated with those for any other bins, i.e., $\sigma_{\rm ex}^2 = \delta^{}_{ij} u^{}_i u^{}_j$.

In addition, two types of theoretical errors in the physics input are taken into account. First, the flavor mixing angle $\theta^{}_{12}$ will be very precisely measured by JUNO, which is able to achieve an unprecedented precision about $0.3\%$ for $\sin^2 \theta^{}_{12}$ by around 2026 \cite{An:2015jdp, Djurcic:2015vqa}. The influence of such a small error on the ultimate sensitivity of the DARWIN-like experiment should be negligible. Second, only two main components of low-energy solar neutrinos are relevant, namely, the $pp$ neutrinos and the ${\rm {}^7\!Be}$ neutrinos, which account for about $91\%$ and $7\%$ of the total solar neutrino flux, respectively. The theoretical errors for the $pp$ and ${\rm {}^7\!Be}$ neutrino fluxes are estimated in the standard solar model to be $0.5\%$ and $6\%$~\cite{Vinyoles:2016djt}, where an anti-correlation between the flux errors of $pp$ and ${\rm {}^7\!Be}$ neutrinos can be observed. In our analysis, this anti-correlation is ignored. Since the anti-correlation could help us to reduce the theoretical error of the total neutrino flux, the results obtained by ignoring it is somehow more conservative. For comparison, we shall present both results with and without the aforementioned theoretical errors.

\begin{figure}[t!]
    \begin{center}
    \hspace{-0.2cm}
    \subfigure{%
    \includegraphics[width=0.47\textwidth]{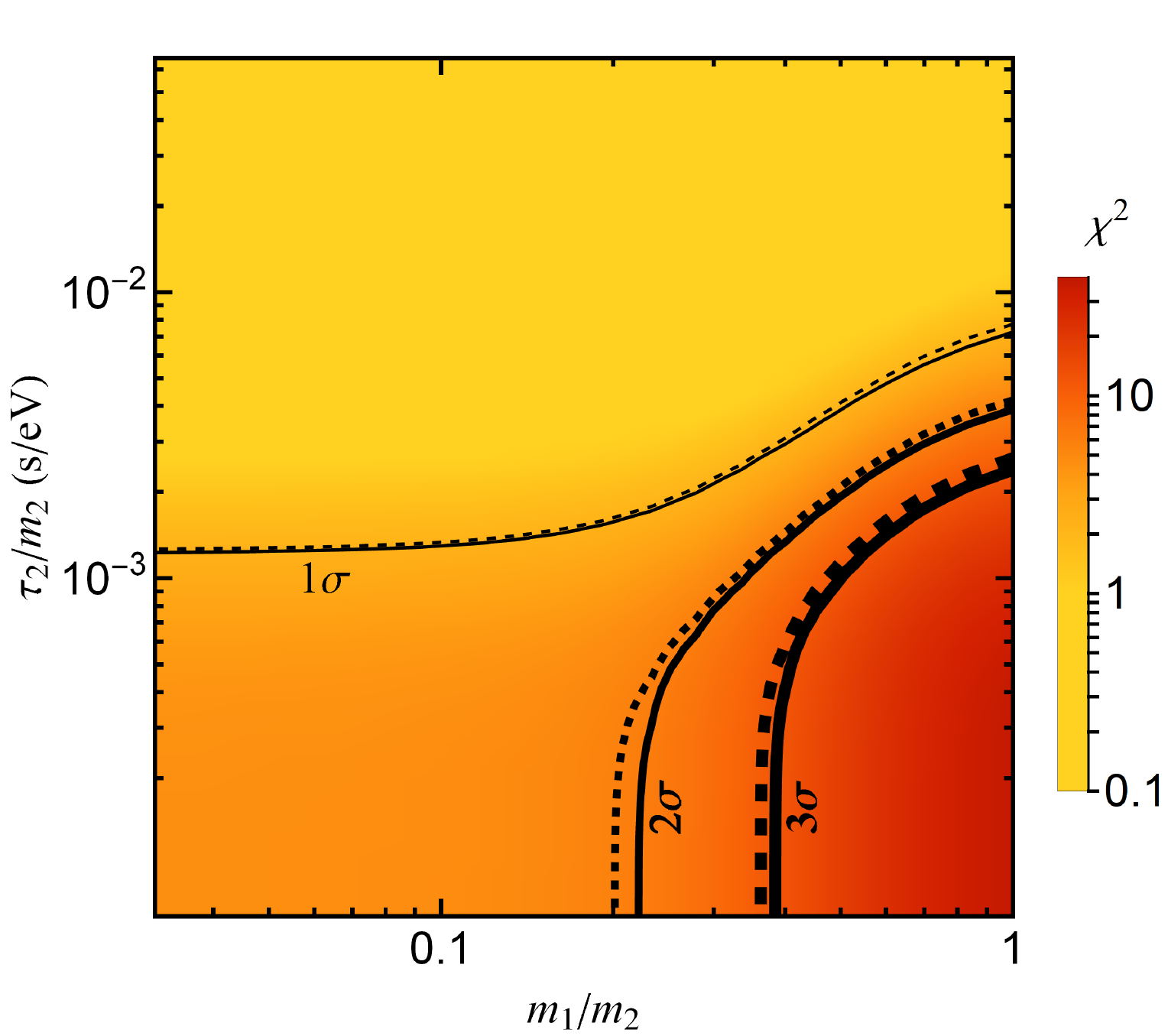}
    }%
    \subfigure{%
    \hspace{0.2cm}
    \includegraphics[width=0.48\textwidth]{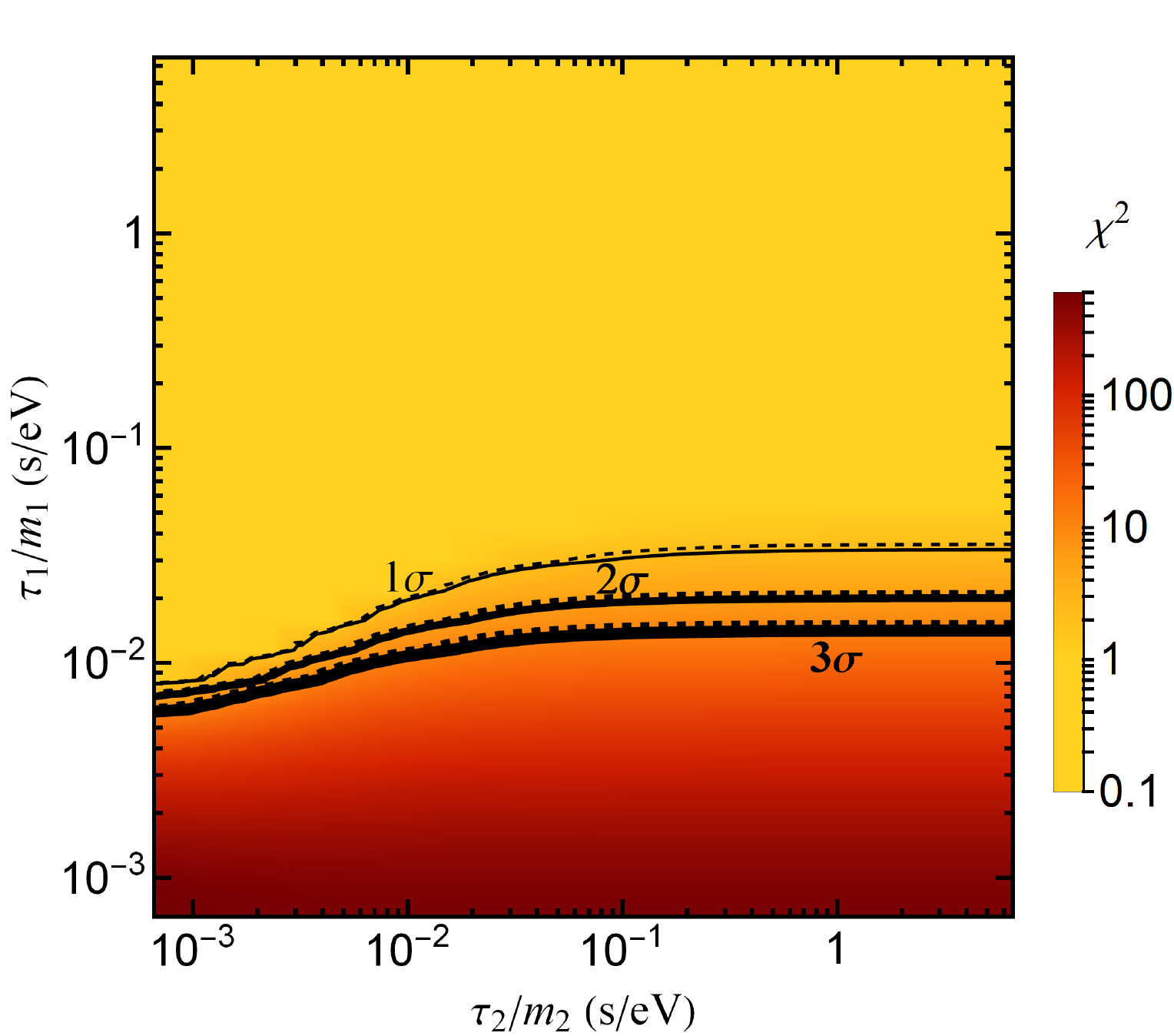}        }
    \end{center}
    \vspace{-0.5cm}
    \caption{The expected sensitivities of a DARWIN-like xenon experiment to the neutrino lifetime $\tau^{}_2/m^{}_2$ with respect to the mass ratio $m^{}_1/m^{}_2$ in the non-radiative visible decay case with normal neutrino mass ordering ({\it Left Panel}), and to $(\tau^{}_1/m^{}_1, \tau^{}_2/m^{}_2)$ in the same case but with inverted neutrino mass ordering ({\it Right Panel}). The experimental setup and the conventions for the contours are the same as those in Fig.~\ref{fig:sen1}, and the branching ratio $f$ for $\nu^{}_2 \to \nu^{}_1$ decays has been marginalized over in the case of inverted mass ordering.}
    \label{fig:sen2}
\end{figure}

In Fig.~\ref{fig:sen1}, the expected sensitivities to the neutrino lifetimes $(\tau^{}_1/m^{}_1, \tau^{}_2/m^{}_2)$ have been shown in the case of IVDs. The exposure of the xenon detector is $70~{\rm ton} \cdot {\rm year}$. Since neutrino mass eigenstates $\nu^{}_1$ and $\nu^{}_2$ will simply disappear, there should be a degeneracy in the contributions from $\nu^{}_1$ and $\nu^{}_2$ decays to the reduction of the ER events at the detector. Consequently, the allowed parameter space is approximately symmetric between $\tau^{}_1/m^{}_1$ and $\tau^{}_2/m^{}_2$. At the $2\sigma$ level, the constraints turn out to be $\tau^{}_1/m^{}_1 \gtrsim 3\times 10^{-2}~{\rm s}/{\rm eV}$ and $\tau^{}_2/m^{}_2 \gtrsim 8\times 10^{-3}~{\rm s}/{\rm eV}$. The difference between these two constraints can be mainly ascribed to the fact $\cos^2 \theta^{}_{12} \approx 2 \sin^2 \theta^{}_{12}$ for $\theta^{}_{12} \approx 34^\circ$. The contours for the case of ignoring theoretical errors in the mixing angle and the initial neutrino fluxes are represented by the dashed curves, which almost overlap with the solid curves for the results including those errors.

In Ref.~\cite{Chen:2016eab}, it has been suggested that the cross section of the elastic neutrino-electron scattering could be reduced by about $23\%$ for the bound-state electrons in the xenon atoms, compared to that for free electrons. If such a change in the cross section is taken into account, we find that the $\chi^2$ values will in general be decreased by about $65\%$. The lower bounds derived in the previous analysis are reduced only by a factor of two, i.e., $\tau^{}_1/m^{}_1 \gtrsim 1.5\times 10^{-2}~{\rm s}/{\rm eV}$ and $\tau^{}_2/m^{}_2 \gtrsim 4\times 10^{-3}~{\rm s}/{\rm eV}$ at the $2\sigma$ level. This conclusion applies as well to the constraints in other scenarios, if the impact of bound-state electrons on the cross section is considered.

In Fig.~\ref{fig:sen2}, we show the exclusion curves for neutrino lifetimes in the case of NRVDs with NO (the left panel) and those with IO (the right panel). In the former case, if the mass ratio is small enough, i.e., $m^{}_1/m^{}_2 \lesssim 0.23$, there is essentially no limit on $\tau^{}_2/m^{}_2$ at the $2\sigma$ level. However, the $1\sigma$ lower bound $\tau^{}_2/m^{}_2 \gtrsim 10^{-3}~{\rm s}/{\rm eV}$ can still be obtained. On the other hand, if the mass spectrum is nearly degenerate, one can see $\tau^{}_2/m^{}_2 \gtrsim 4\times 10^{-3}~{\rm s}/{\rm eV}$ at the $2\sigma$ level. As we have emphasized before, the coupling constant between neutrinos and the massless particle in the hidden sector should be very large in order to compensate for the tiny phase space in the limit of $m^{}_1/m^{}_2 \to 1$. So the experimental bound on the coupling will be extremely weak.

The sensitivities to $(\tau^{}_1/m^{}_1, \tau^{}_2/m^{}_2)$ in the IO case have been presented in the right panel of Fig.~\ref{fig:sen2}. In our statistical analysis, the branching fraction $f$ for $\nu^{}_2 \to \nu^{}_1$ decays is marginalized over its allowed range of $[0, 1]$ to minimize the $\chi^2$ value. At the $2\sigma$ level, we observe a lower bound $\tau^{}_{1}/m^{}_1 \gtrsim 8\times 10^{-3}~{\rm s}/{\rm eV}$ for a short lifetime of $\nu^{}_2$ (e.g., $\tau^{}_2/m^{}_2 \lesssim 10^{-3}~{\rm s}/{\rm eV}$), but the bound becomes slightly tighter, namely, $\tau^{}_{1}/m^{}_1 \gtrsim 2\times 10^{-2}~{\rm s}/{\rm eV}$, for a rather stable $\nu^{}_2$ (e.g., $\tau^{}_2/m^{}_2 \gtrsim 1~{\rm s}/{\rm eV}$). This feature can be understood by noticing that the branching ratio $f$ can always be optimized to balance the increase in the event number from the $\nu^{}_1$ production and the decrease from the invisible decays into $\nu^{}_3$.

A useful remark should be made on the assumption of a stable $\nu^{}_3$ in the previous discussions. This will simplify the decay patterns of neutrinos, and the statistical analysis as well by reducing the number of fitting parameters, namely, $\tau^{}_3/m^{}_3$ and the branching ratio for different decay channels. To estimate the impact of relaxing this assumption, we have made a comparison with the scenario of $\nu^{}_3$'s decaying into $\nu^{}_1$'s, which is relevant for our discussions about the NRVDs with the NO. It turns out that the change in the $\chi^2$ value is only about $0.6$, which is comparable to the difference between the solid and dashed curves in Fig.~\ref{fig:sen1} and Fig.~\ref{fig:sen2}. Finally, it is worthwhile to mention that the particle $\eta$ in the NRVD case is assumed to be massless for simplicity. As we have demonstrated for $\nu^{}_2 \to \nu^{}_1 + \eta$ decays, the mass ratio $m^{}_1/m^{}_2$ plays an important role. For the same reason, the ratio of $m^{}_\eta$ to absolute neutrino masses $m^{}_i$ should also be taken into full consideration when a concrete model of neutrino decays and the interactions beyond the SM are given.
\section{Constraints on Neutrino Magnetic Moments}
    \begin{figure}[t!]
    \begin{center}
    \hspace{-0.2cm}
    \subfigure{%
    \includegraphics[width=0.47\textwidth]{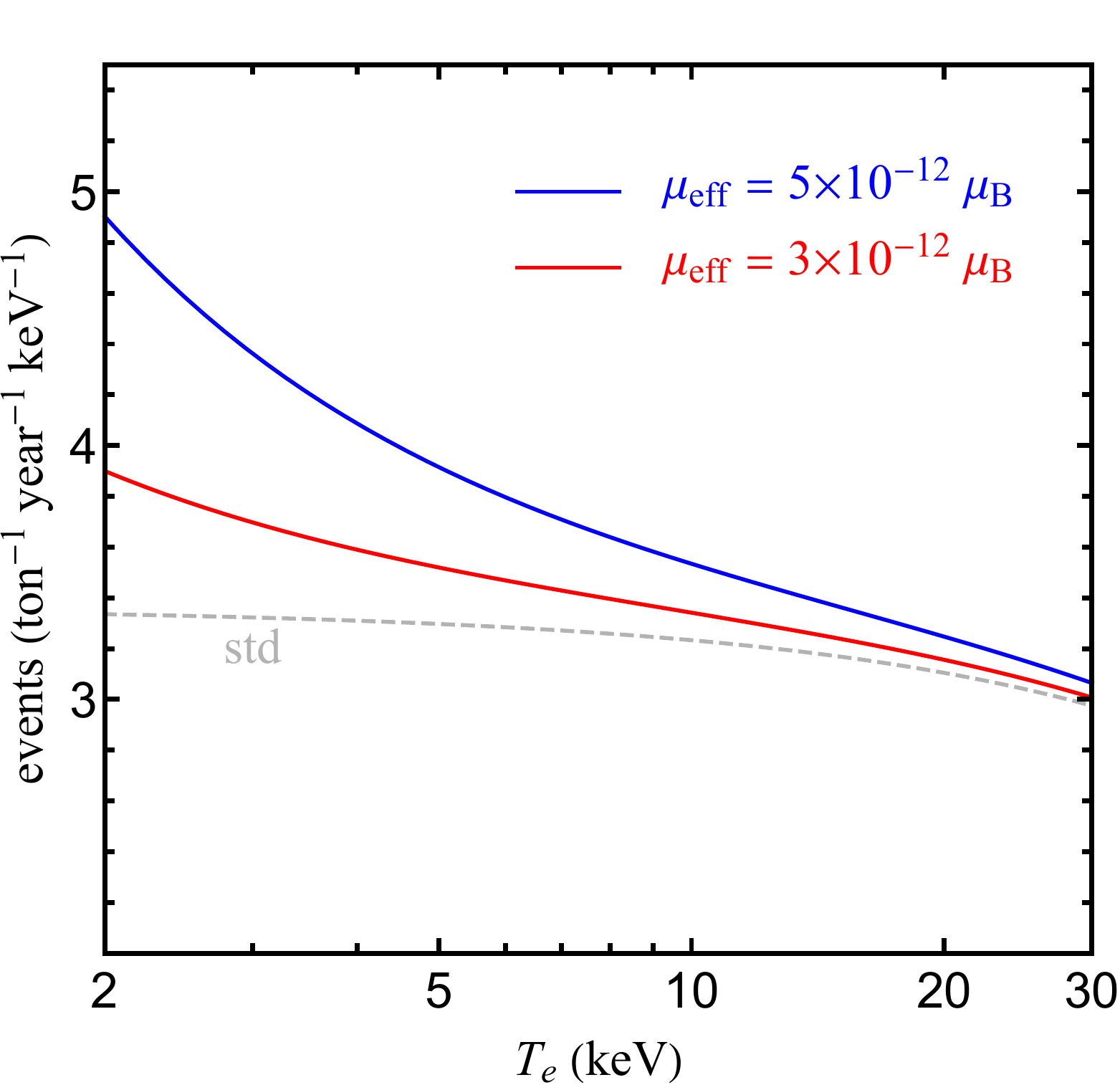}
    }%
    \end{center}
    \vspace{-0.5cm}
    \caption{Illustrations for the event rates of the elastic neutrino-electron scattering in the DARWIN-like experiment for two different values of the neutrino magnetic moment $\mu^{}_{\rm eff} = 3 \times 10^{-12}~\mu^{}_{\rm B}$ and $\mu^{}_{\rm eff} = 5 \times 10^{-12}~\mu_{\rm B}$, where the event rate in the SM has been shown as the dashed curve.}
    \label{fig:event5}
    \end{figure}

As we have already stressed in the introduction, the RDs of massive neutrinos are intimately related to their electromagnetic properties. Therefore, the investigation of experimental constraints on neutrino magnetic moments in this section can be viewed as an independent and complementary study of the neutrino RDs. Current limits on neutrino magnetic moments have been derived, based on a variety of observations from astrophysics, cosmology and particle-physics experiments in terrestrial laboratories~\cite{Raffelt:1999gv, Giunti:2015gga}.

First, if massive neutrinos are Dirac particles, the right-handed neutrino components would be efficiently produced in the primordial magnetic fields~\cite{Enqvist:1994mb, Long:2015cza, Zhang:2015wua}. These extra components contribute to the effective degrees of freedom in radiation, which should be severely constrained by the Big Bang Nucleosynthesis~\cite{Tanabashi:2018oca} and CMB~\cite{Ade:2015xua} observations. For instance, for the primordial magnetic field $B^{}_0 = 10^{-14}~{\rm G}$ and the size of the magnetic field domain $\lambda = 1~{\rm Mpc}$~\cite{Giunti:2015gga}, the upper bound $\mu^{}_{\rm eff} \lesssim 10^{-16}~\mu_{\rm B} $ can be derived. Second, the plasmon decaying into neutrinos in the globular clusters can be implemented to set a $1\sigma$ limit of $\mu^{}_{\rm eff} \lesssim (2.2 \cdots 2.6) \times 10^{-12}~\mu_{\rm B} $~\cite{Raffelt:1999gv,Arceo-Diaz:2015pva,Viaux:2013hca,Viaux:2013lha}, which is as stringent as that from the neutrino observations of Supernova 1987A~\cite{Barbieri:1988nh}. Finally, the GEMMA experiment~\cite{Beda:2012zz} yields the limit $\mu^{}_{\rm eff}  \lesssim 2.9 \times 10^{-11}~\mu^{}_{\rm B}$ by observing the low-energy ER events caused by reactor neutrinos, and the future GEMMA-III project may improve the sensitivity to $\mu^{}_{\rm eff} \lesssim 9\times 10^{-12}~\mu^{}_{\rm B}$ by further lowering down the energy threshold and shortening the distance between the reactor and the detector~\cite{Giunti:2014ixa}.

The Borexino observation of low-energy solar neutrinos places a useful constraint on the neutrino magnetic moment~\cite{Borexino:2017fbd, Canas:2015yoa, Barranco:2017zeq, Khan:2017djo,Guzzo:2012rf}, while the most recent bound~\cite{Borexino:2017fbd, Khan:2017djo} has been updated with the Borexino-II data~\cite{Agostini:2017ixy}, namely, $\mu^{}_{\rm eff} \lesssim 10^{-11}~\mu^{}_{\rm B} $ at the $90\%~{\rm C.L.}$ As is well known, neutrinos with lower energies will be advantageous in constraining neutrino lifetimes and magnetic moments, so the experiments for the direct detection of dark-matter particles are suitable for this task. In the literature, the dark-matter experiments have been implemented to probe non-standard neutrino interactions~\cite{Dutta:2017nht, Khan:2017oxw, Cerdeno:2016sfi, Billard:2014yka, AristizabalSierra:2017joc, Harnik:2012ni,Papoulias:2018uzy}. It is worth mentioning that Ref.~\cite{Harnik:2012ni,Papoulias:2018uzy} have explored the possibility to examine neutrino magnetic moments via the NR events in the xenon detector, e.g., Fig.~4 in Ref.~\cite{Harnik:2012ni} and Fig.~11 in Ref.~\cite{Papoulias:2018uzy}. However, the sensitivity will be competitive only if the threshold of the NR energy is as low as $10^{-3}~{\rm keV}$, which is far below the typical value at present.
\begin{figure}[t!]
    \begin{center}
    \subfigure{%
    \includegraphics[width=0.43\textwidth]{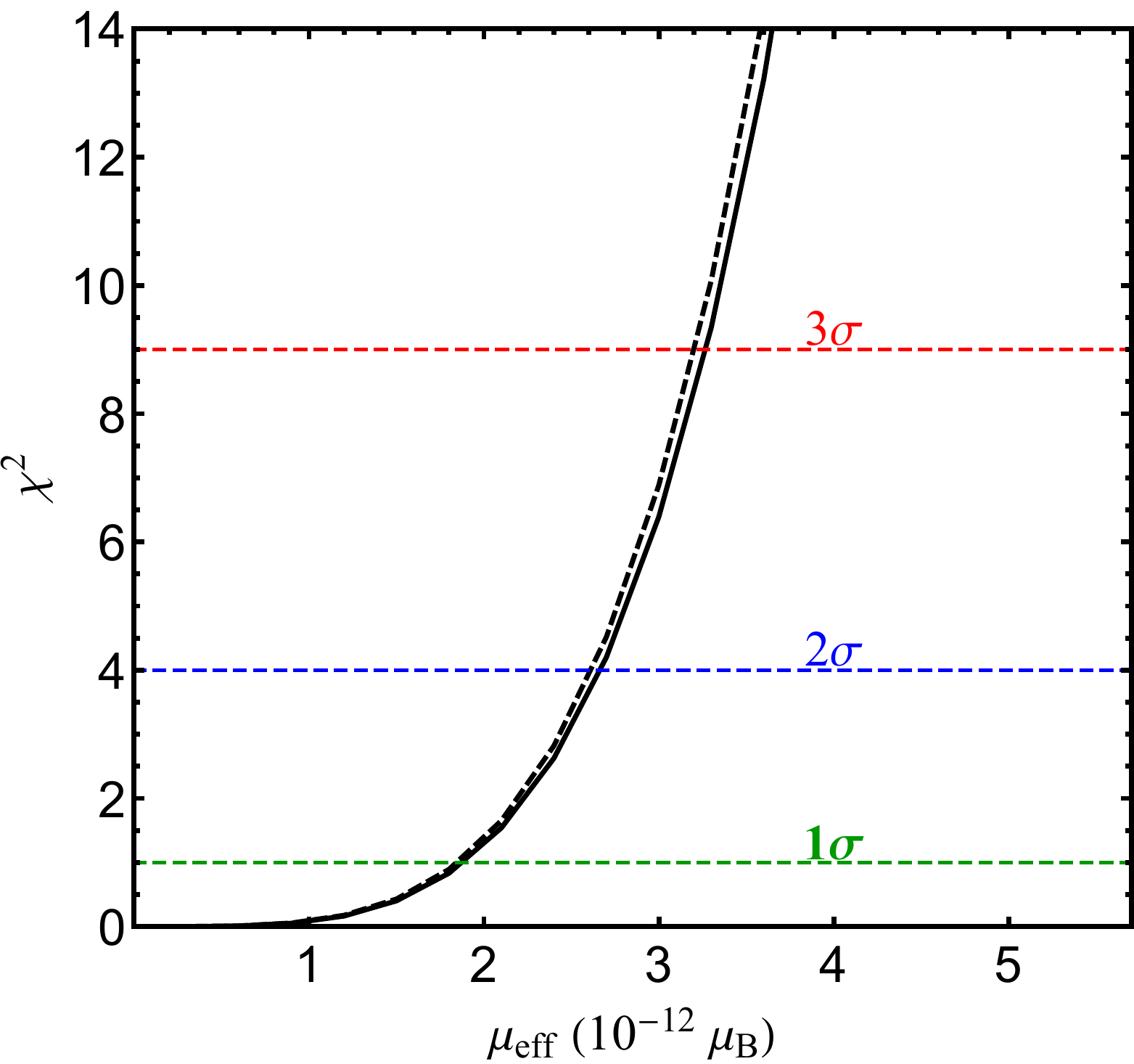}        }%
    \subfigure{%
    \hspace{0.5cm}
    \includegraphics[width=0.432\textwidth]{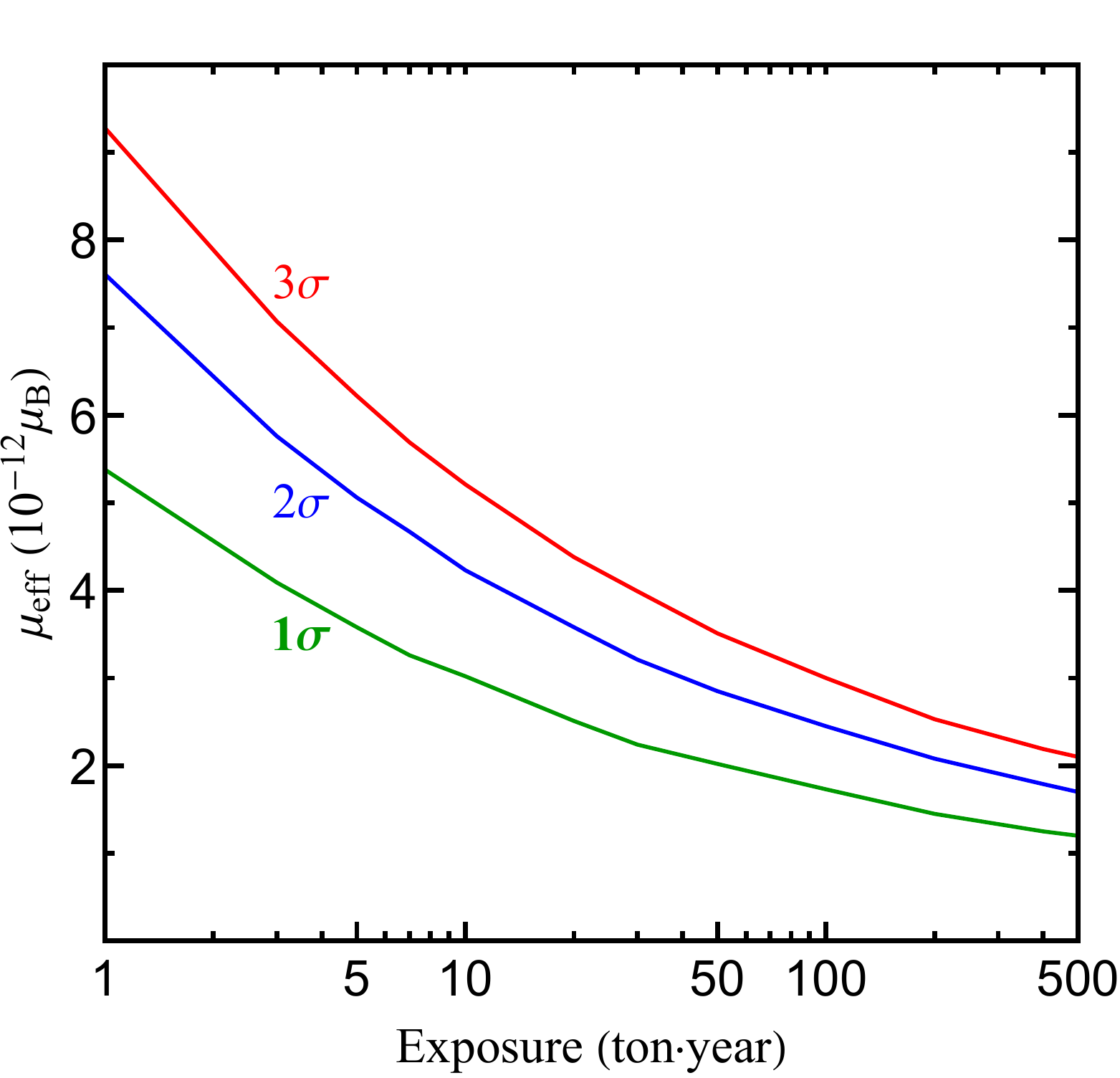}        }
    \end{center}
    \vspace{-0.5cm}
    \caption{The $\chi^2$ distribution for the effective neutrino magnetic moment $\mu^{}_{\rm eff}$ ({\it Left Panel}), which is supposed to be responsible for the ER events in a DARWIN-like experiment, and the future $1\sigma$, $2\sigma$ and $3\sigma$ upper bounds with respect to the total exposure ({\it Right Panel}). The solid (dashed) curve in the left panel stands for the $\chi^2$ with (without) theoretical errors.}
    \label{fig:sen4}
\end{figure}

Now suppose that the ER events in the xenon detector are induced by the solar neutrinos via both weak and electromagnetic interactions. In particular, the contributions from neutrino magnetic moments will be dominant at low energies. The differential cross section reads~\cite{Giunti:2014ixa}
\begin{align}
\frac{{\rm d} \sigma^{}_{\rm m}(T^{}_e, E^{}_{\nu})}{{\rm d} T^{}_e} = \frac{\pi \alpha^2}{m_e^2} \left(\frac{\mu^{}_{\rm eff}}{\mu^{}_{\rm B}}\right)^2\left(\frac{1}{T_e} - \frac{1}{E_{\nu}}\right) \; ,
\end{align}
where $\alpha$ is the fine-structure constant and $\mu_{\rm eff}$ is the effective neutrino magnetic moment, which has been defined as follows
\begin{align} \label{eq:munew}
\mu_{\rm eff}^2 = \sum_{ij}|U_{ei}|^2 \left(|\mu^{}_{ij}|^2 + |\epsilon^{}_{ij}|^2\right) \; .
\end{align}
Note that the definition of $\mu^{}_{\rm eff}$ in Eq.~(\ref{eq:munew}) is different from that in Eq.~(\ref{eq:decayrate}). These definitions will become almost identical when the transitional electric and magnetic dipole moments are on the same order of magnitude and the unitarity condition $|U^{}_{e1}|^2 + |U^{}_{e2}|^2 + |U^{}_{e3}|^2 = 1$ is satisfied. The complete determination of all the individual electric and magnetic dipole moments will be very challenging, so we follow the definition of $\mu^{}_{\rm eff}$ in Eq.~(\ref{eq:munew}) and try to extract the experimental bound on it in this section.

In Fig.~\ref{fig:event5}, the event spectra in a DARWIN-like experiment have been given for two different values of the neutrino magnetic moment, namely, $\mu^{}_{\rm eff} = 3\times 10^{-12}~\mu^{}_{\rm B}$ and $\mu^{}_{\rm eff} = 5\times 10^{-12}~\mu^{}_{\rm B}$. As we expect, the events rate is larger for lower recoil energies, and the deviation from that with the standard interaction can be as large as $15\%$ in the case of $\mu^{}_{\rm eff} = 3\times 10^{-12}~\mu^{}_{\rm B}$.

Following the similar procedure for a statistical analysis in the previous section, we attempt to constrain neutrino magnetic moments. The future sensitivity has been plotted in left panel of Fig.~\ref{fig:sen4}, where one can observe that the $2\sigma$ upper bound is $\mu^{}_{\rm eff} \lesssim 2.6 \times 10^{-12}~\mu^{}_{\rm B} $, which means one order of magnitude improvement of the Borexino constraint and is comparable to the astrophysical bound. Although this is already among the best results that will be available in the near future, it seems quite difficult to further increase the experimental sensitivity. For example, as shown in the right panel of Fig.~\ref{fig:sen4}, even for a double exposure of $140~{\rm ton}\cdot{\rm year}$, the upper limit would only be slightly improved to be $\mu^{}_{\rm eff}\lesssim 2.3 \times 10^{-12}~\mu^{}_{\rm B} $.

\section{Summary}

Motivated by the recent tremendous progress in the direct searches for dark-matter particles, we have performed a sensitivity study of the future large xenon detectors to the intrinsic properties of massive neutrinos, such as neutrino lifetimes and magnetic moments, via the detection of low-energy solar neutrinos.

For a DARWIN-like experiment, which is capable of observing the electron-recoil events in the energy window of $(2 \cdots 30)~{\rm keV}$, we derive the expected experimental constraints on the parameters $\tau^{}_i/m^{}_i$ in different scenarios of neutrino decays. For the invisible decays, the lower bounds $\tau^{}_1/m^{}_1 \gtrsim 3 \times 10^{-2}~{\rm s}/{\rm eV}$ and $\tau^{}_2/m^{}_2 \gtrsim 8\times 10^{-3}~{\rm s}/{\rm eV}$ at the $2\sigma$ level can be obtained. These limits are better by about one order of magnitude than those derived from current neutrino oscillation experiments including Borexino, KamLAND and DayaBay. For the non-radiative visible decays, we have emphasized the important role played by the unknown neutrino mass spectrum. If the neutrino mass spectrum is hierarchical in the case of normal mass ordering, the constraints on neutrino lifetimes are very weak. If the mass spectrum is nearly degenerate, the bound $\tau^{}_{2}/m^{}_2 \gtrsim 3\times 10^{-3}~{\rm s}/{\rm eV}$ at the $2\sigma$ C.L. can be derived. For the inverted mass ordering, only the $\nu^{}_1$ decays can be actually constrained, i.e., $\tau^{}_1/m^{}_1 \gtrsim 10^{-2}~{\rm s}/{\rm eV}$. Finally, for a total exposure of $70~{\rm ton} \cdot {\rm year}$, the $2\sigma$ upper bound on the neutrino magnetic moment has been found to be $\mu^{}_{\rm eff} \lesssim 2.6\times 10^{-12} ~\mu^{}_{\rm B}$. This is among the best limits that can be obtained from the laboratory experiments in the near future.

The detection of low-energy neutrinos is one of the most important directions in neutrino physics. As for solar neutrinos, the thermal fluxes around the keV energies have been calculated in Ref.~\cite{Vitagliano:2017odj}, and the final detection of these neutrinos should be very helpful in exploring the intrinsic properties of massive neutrinos. In this connection, the developments in the dark-matter experiments, as well as in the atomic physics and quantum optics, will hopefully provide us with new possibilities soon.

\section*{Acknowledgements}

This work was supported in part by the National Natural Science Foundation of China under Grant No. 11775232 and No. 11835013, and by the CAS Center for Excellence in Particle Physics (CCEPP).

\section*{Appendix A: Derivation of Eq.~(\ref{eq:IOprob})}
For the NRVDs in the IO case, there are three relevant decay processes, i.e., $\nu^{}_2 \to \nu^{}_1 + \eta$, $\nu^{}_2 \to \nu^{}_3 + \eta$, and $\nu^{}_1 \to \nu^{}_3 + \eta$. As has been mentioned before, the masses of $\nu^{}_1$ and $\nu^{}_2$ are nearly degenerate and thus the daughter neutrino will have an energy very close to the mother neutrino energy for the decay process $\nu^{}_2 \to \nu^{}_1 + \eta$. However, the mass ratio of $m^{}_3/m^{}_1$ or $m^{}_3/m^{}_2$ can be either small or close to one. For definiteness, we assume that the neutrino mass spectrum is nearly degenerate, i.e., $m^{}_1 \sim m^{}_2 \sim m^{}_3$. In this case, the energy distribution of the daughter neutrino is peaked around the mother neutrino energy for all three decay processes. Under such an assumption, we can write down the evolution equations of the neutrino fluxes
\begin{align} \label{eq:16}
\frac{\mathrm{d}}{\mathrm{d}t} \left( \frac{\mathrm{d}\phi^{}_{1}}{\mathrm{d}E}\right) = & -\frac{\mathrm{d}\phi^{}_{1}}{\mathrm{d}E} \cdot \frac{m^{}_{1}}{\tau^{}_{1}E }
+\frac{\mathrm{d}\phi^{}_{2}}{\mathrm{d}E} \cdot \frac{m^{}_{2}f}{\tau^{}_{2}E }\;,  \\
\label{eq:17}
\frac{\mathrm{d}}{\mathrm{d}t} \left( \frac{\mathrm{d}\phi^{}_{2}}{\mathrm{d}E}\right) = & -\frac{\mathrm{d}\phi^{}_{2}}{\mathrm{d}E} \cdot \frac{m^{}_{2}}{\tau^{}_{2}E }\;,  \\
\label{eq:18}
\frac{\mathrm{d}}{\mathrm{d}t} \left( \frac{\mathrm{d}\phi^{}_{3}}{\mathrm{d}E}\right) = & ~ \frac{\mathrm{d}\phi^{}_{1}}{\mathrm{d}E} \cdot \frac{m^{}_{1}}{\tau^{}_{1}E } +
\frac{\mathrm{d}\phi^{}_{2}}{\mathrm{d}E} \cdot \frac{m^{}_{2}(1-f)}{\tau^{}_{2}E }\;,
\end{align}
where $t$ is the flight time of neutrinos, the factor of $m^{}_{i}/E$ (for $i=1,2$) arises from the time dilation effect in the laboratory frame, and $f$ denotes the branching fraction of $\nu^{}_2 \to \nu^{}_1$ decays. The total probability is conserved due to
\begin{align} \label{eq:probconser}
\frac{\mathrm{d}}{\mathrm{d}t} \left( \frac{\mathrm{d}\phi^{}_{1}}{\mathrm{d}E} +\frac{\mathrm{d}\phi^{}_{2}}{\mathrm{d}E} +\frac{\mathrm{d}\phi^{}_{3}}{\mathrm{d}E}\right) = 0 \;.
\end{align}
Using the initial condition of ${\mathrm{d}\phi^{0}_{i}}/{\mathrm{d}E} = |U^{}_{ei}|^2 \cdot {\mathrm{d}\phi }/{\mathrm{d}E}$, we can solve the neutrino fluxes at any flight time by integrating the above differential equations. More explicitly, one can first integrate Eq.~(\ref{eq:17}) to easily get the expression of ${\mathrm{d}\phi^{}_{2}}/{\mathrm{d}E}$ and then substitute it into Eq.~(\ref{eq:16}) to obtain ${\mathrm{d}\phi^{}_{1}}/{\mathrm{d}E}$. The result of ${\mathrm{d}\phi^{}_{3}}/{\mathrm{d}E}$ can be simply derived from ${\mathrm{d}\phi^{}_{1}}/{\mathrm{d}E}$ and ${\mathrm{d}\phi^{}_{2}}/{\mathrm{d}E}$ with the help of the probability conservation in Eq.~(\ref{eq:probconser}). Dividing the final results by the original flux ${\mathrm{d}\phi }/{\mathrm{d}E}$, we can then obtain
those expressions in Eq.~(\ref{eq:IOprob}).


\end{document}